\documentclass[11pt]{amsart}
\usepackage{amsaddr}

\usepackage{amsmath,amsthm,amsfonts}
\usepackage{tensor}
\usepackage{graphicx}
\usepackage{color}
\usepackage{xcolor}
\usepackage{tikz}
\usepackage{float}
\usepackage[margin=2cm]{geometry}
\usepackage{bbm}
\usepackage[lofdepth,lotdepth]{subfig}

\usepackage{ulem}
\usepackage{verbatim}

\newtheorem{prop}{Proposition}[section]

\newcommand{\beprop}{\begin{prop}}
\newcommand{\enprop}{\end{prop}}
\newcommand{\bprf}{\begin{proof}}
\newcommand{\eprf}{\end{proof}}

\usepackage[default]{frcursive}
\usepackage[T1]{fontenc}

\usepackage{color}
\usepackage{xcolor}
\usepackage[breaklinks,colorlinks,backref]{hyperref}
\hypersetup{
    colorlinks, 
    linktoc=all, 
    linkcolor=red,  
  citecolor=hyptxt,
  urlcolor=blue}
  \hypersetup{
  citebordercolor=,
  filebordercolor=green,
  linkbordercolor=blue
}
\definecolor{hyptxt}{rgb}{0.7, 0.4, 0.9}
\usepackage{bibtopic}

\usepackage[full]{textcomp} 
     \usepackage{fbb} 
     \usepackage[scaled=.95]{cabin}
     \usepackage[varqu,varl]{inconsolata}
     \usepackage[libertine,bigdelims,vvarbb]{newtxmath}
     \usepackage[cal=boondoxo]{mathalfa}
     \usepackage[T1]{fontenc}
\useosf

\definecolor{hervecolor}{rgb}{0.8,0,0.7}
     
\newcommand{\ket}[1]{|\kern.3ex#1\kern.3ex\rangle}
\newcommand{\bra}[1]{\langle\kern.3ex #1 \kern.3ex|}
\newcommand{\scalar}[2]{\langle\kern.3ex #1 \kern.3ex|\kern.3ex#2\kern.3ex\rangle}

\newcommand{\NN}{\mathcal{N}}

\newcommand{\ii}{\mathsf{i}}

\def\R{\mathbb{R}}
\def\N{\mathbb{N}}
\def\C{\mathbb{C}}

\def\lg{\langle }
\def\rg{\rangle }
\def\deq{\stackrel{\mathrm{def}}{=}}

\def\vk{\varkappa}

\def\ud{\mathrm{d}}

\def\sfP{\mathsf{P}}

\def\id{\mathbb{1}}

\def\hN{\hat N}

\def\bsh{\boldsymbol{\mathsf{h}}}

\numberwithin{equation}{section}

\begin{document}
\title{Lowering Helstrom Bound with non-standard coherent states}
\author{\small Evaldo M. F. Curado${}^{a,b}$, Sofiane Faci${}^{a, c}$,  Jean-Pierre Gazeau${}^{a,d}$, and Diego Noguera${}^a$}
\address{${}^a$ Centro Brasileiro de Pesquisas F\'{\i}sicas, Rua Xavier Sigaud 150,  Rio de Janeiro, Brazil}
\address{${}^b$  National Institute of Science and Technology for Complex Systems,  Rua Xavier Sigaud 150, Rio de Janeiro, Brazil } 
\address{${}^c$ Universidade Federal Fluminense, RCN/IHS, Rio das Ostras, Brazil}
\address{${}^d$ Universit\'e de Paris, CNRS, Astroparticule et Cosmologie, F-75013 Paris, France}, 
\email{evaldo@cbpf.br, sofiane@cbpf.br, gazeau@apc.in2p3.fr, diegomac@cbpf.br}
\begin{abstract}
In quantum information processing, {using a receiver device to differentiate between two nonorthogonal states leads to a quantum error probability. The minimum possible error is} known as the Helstrom bound.  In this work we study and compare  quantum limits for states which generalize the Glauber-Sudarshan coherent states, like non-linear, Perelomov, Barut-Girardello, and (modified)  Susskind-Glogower coherent states. For some of these,   we show that  the  Helstrom bound can be significantly lowered and  even vanish in specific regimes. 
\end{abstract}

\maketitle
\tableofcontents

\section{Introduction}
\label{intro}
In quantum information processing, {quantum operations represent communication channels while
quantum states are the information carriers. The sender encodes information} into a {state $\rho$ which pertains to an alphabet  $\mathcal{A}=\{\rho_0, \rho_1, \dotsc, \rho_M\}$. 
To find out which state was sent, the receiver carries out a measurement.}
{ When the transmitted states  are not orthogonal,}  errors are possible and { a nonzero probability exists that the receiver} misconstrues { the transmitted information} (see \cite{holevo73} \cite{main:peres}  \cite{main:ch4:fuchsphd} \cite{holevo11} and references therein). Note that the impossibility of differentiating between 
{ nonorthogonal states  represents an advantage when dealing  with quantum
key distribution \cite{main:ch4:benbra}. In fact, using a nonorthogonal alphabet implies that any measurement disturbs the state and thus allows to impede an eavesdropper from obtaining information without being noticed.
Moreover, as  shown by Fuchs \cite{main:ch4:fuchsphd}, in some setups, nonorthogonality  actually maximizes the classical information capacity in noisy channels. }
 
{ In order to differentiate between nonorthogonal states one can optimize a state-determining measure over all POVMs, or positive operator valued measures.} In \cite{holevo73}, the author  gives a systematic treatment of quantum statistical decision theory based on POVM. Necessary and sufficient conditions are given for optimality. {Moreover the notion of  maximum-likelihood measurement is introduced.} 
  
{When trying to minimize the error in measurement over all possible POVM's we are led to the  \textit{quantum error probability}, also known as \textit{Helstrom bound} \cite{main:helstrom}. This limit leads to a criterium of quality when discriminating between the transmitted nonorthogonal states. }

For an on-off keyed alphabet (the pair formed by the Fock vacuum and an arbitrary coherent state) where the state discrimination is based on direct photon-counting, a discrimination error can occur when no photons are detected due to vacuum fluctuations. The discrimination error probability in this case is called the shot-noise-error probability (sometimes called standard quantum limit), it is always higher than the Helstron bound and is inversely proportional to the exponential of the average number of photons. Surpassing the shot noise in order to achieve the Hesltrom bound is an important subject of study in numerous works related to optical coherent state discrimination \cite{main:ch4:comage,tsujino11,becerra13,Kunz:2019,Sych:2016,DiMario:2019}. 
  
{A series of experimental and theoretical works using standard CS, i.e., Glauber-Sudarshan coherent states (or GS-CS) as nonorthogonal states were published, see \cite{main:ch4:comage,gazeaubook09} and references therein, for instance.  On the one hand, GS-CS describe ideal lasers, on the other hand exhibit important mathematical properties. 
These are eigenstates of the  annihilation operator in Fock Hilbert space and yield a \textit{Poissonian} number distribution. 
In fact, real lasers are known to be better represented by states which are almost or non-Poissonian distributions  \cite{perina,perina01}. 
Deviation from Poissonian statistics is also a consequence of non-linear effects in photons production. 
It is therefore important to study the mathematical and physical properties of ``generalized'' coherent states that exhibit both coherent and nonlinear behaviours.}  
We should also mention recent results showing how using non-standard CS greatly enhances sensitivities of gravitational waves detectors, for both LIGO \cite{Tse:2019wcy} and VIRGO \cite{Acernese:2019sbr}.

{The present study concerns  a class  of  non standard coherent states (the so-called ``AN-CS'') which was  introduced in  \cite{gazeau19} and which encompasses most of the known generalisations of the Glauber-Sudarshan CS. They are denoted by  $|\alpha;\bsh\rg$. We  explore} some of their statistical properties in terms of photon detection and  examine   if it is possible to lower   the Helstrom bound in comparison with  the GS-CS. This program was initiated by two of us in the article \cite{cugaro10}, and we intend  to pursue it with the present work. We have judged sufficient to  restrict our study to the simplest case of the overlap $\vert \lg \alpha;\bsh|0\rg\vert$ with the vacuum, since  the GS-CS's themselves exhibit  the well-known  property $\vert\lg\alpha|\alpha^{\prime}\rg\vert= \vert\lg \alpha- \alpha^{\prime}|0\rg\vert$. We show that several non standard CS do indeed allow to decrease and even vanish the Helstrom limit in some regimes (for a given mean photons number), and this is the  main outcome of our study.

In Section \ref{helstrom}  we give a short overview of the Helstrom bound in binary communication based on POVM quantum measurement, and its value when we deal with GS-CS  in the cases of perfect and imperfect detection. 
In Section \ref{ANclass}  {we recall the definition of the above mentioned AN-CS class of non standard coherent states \cite{gazeau19} and  their most relevant statistical features, particularly in terms of the Mandel parameter, needed for implementing the present study devoted to  the Helstrom bound.} 
In Section \ref{HB-AN} we examine the Helstrom bound for binary communication involving the vacuum and  an arbitrary state in the AN-CS class, for both perfect and imperfect detections.  
In Section \ref{HBNLCS} we apply the above formalism to the so-called non-linear coherent states, a particular set of  AN-CS.  We examine  
in Section \ref{manhelnlex} the cases of SU$(2)$ and SU$(1,1)$ coherent states viewed as optical CS and unveil unexpected aspects of these states within the framework of quantum optics. In particular, we consider as quite appealing from a physical point of view our interpretation of SU$(2)$, i.e., spin, coherent states as a natural approximation of the  Glauber-Sudarshan CS when the length of the beam is finite, which is actually always the case. 
We proceed  in Section \ref{defbincs} with a similar study for non-linear CS built from deformed binomial distributions introduced in a series of works by two of us \cite{cugaro11,bercugaro12,bercugaro13}.  
Section \ref{sussglo} is devoted to the study  of the Susskind-Glogower CS (see \cite{moyasoto11} and references therein) and  a modified  version that we introduce in this paper. 
We then conclude in Section \ref{conclu} with a discussion about experimental and further theoretical possibilities offered by our results. 

\section{Hestrom Bound in binary communication}
\label{helstrom}
\subsection{Definition}
{ Let us consider a sender using nonorthogonal states as codewords, while the
receiver executes a quantum measurement, $M$, on the channel in order to find out which state was sent.
The measurement $M$ is set through a POVM defined by a
 complete (i.e., countable) set of positive operators \cite{main:peres} that do resolve the identity,}
\begin{equation}
    \sum_i M_i = \id\, \quad
    M_i \geq 0\, ,
\end{equation}
{where $i$ stands for all possible outcomes. }
 
{ Dealing with binary communication, there are two quantum measurement, $M_0, M_1$ and the POVM resolution of unity then reads $M_0 + M_1 = \id$.
The receiver can choose among two hypotheses: i) the transmitted state is $\rho_0$ for when the measurement outcome correlates with $M_0$, ii) the transmitted state is $\rho_1$, if the outcome corresponds to $M_1$.}
 
{An error might occur when there is a  possibility for the receiver to
measure one state while the sender actually had sent the other state. With binary communication there are two possible errors, these go with the following conditional probabilities:
\begin{equation}
\label{errorprob1}
   p(M_0 | \rho_1) = \mathrm{tr}[M_0 \rho_1]
   = \mathrm{tr}[(I- M_1) \rho_1 ]\,, \quad  p(M_1 | \rho_0) =
       \mathrm{tr}[M_1 \rho_0] \,.
\end{equation}
Now, the total  error probability reads
\begin{equation}
\label{errorprob2}
    p[M_0,M_1] =
    \xi_0 p({M}_1| \rho_0) + \xi_1 p({M}_0|\rho_1) \,, \quad \xi_0 + \xi_1 = 1\, , 
\end{equation}
with $\xi_0=p(\rho_0)$ and $\xi_1 = p(\rho_1)$ standing for
the classical probabilities that the sender might send $\rho_0$ and
$\rho_1$, respectively; they encode the prior knowledge of the receiver about the sender chosen states.
In many cases one uses $\xi_0= \xi_1 = 1/2$.}
  
{Minimizing the receiver measurement error over $M_0$ and $M_1$ yields the \textit{Helstrom bound}, or \textit{quantum error probability},
\begin{equation}
  P_\mathrm{H} \equiv \min_{M_0,M_1}
    p[M_0,M_1] \,.
\end{equation}
This is the smallest physically permitted error probability, given that the states $\rho_0$ and $\rho_1$ overlap.}
 
%

{Using pure states, $\rho_0=|\Psi_0\rangle\langle\Psi_0|$ and $\rho_1=|\Psi_1\rangle\langle\Psi_1|$, 
the Helstrom bound can be written as, 
\begin{equation} \label{Equation::PerfectHelstrom}
  P_\mathrm{H} = 
  \frac{1}{2} \left(1-\sqrt{1-4 \xi_0 \xi_1 | \langle \Psi_1 |
  \Psi_0 \rangle |^2} \right) \,.
\end{equation}
The details can be found in \cite{gazeau19}, for instance. Notice that, as would be expected, the quantum error vanishes for orthogonal states, i.e., $ \langle \Psi_1 |  \Psi_0 \rangle=0$.}

\subsection{Helstrom Bound with Glauber-Sudarshan Coherent States}
\subsubsection{Perfect detection}
Light beams can be used as carriers of information in quantum communication.
The Glauber-Sudarshan Coherent States (GS-CS) (also called linear or standard) are defined as the superposition of photon number states $|n\rangle$ given by
\begin{equation}
\label{canonCS1}
|\alpha\rg:=e^{-\frac{|\alpha|^{2}}{2}}\sum_{n=0}^{\infty}\frac{\alpha^n}{\sqrt{n!}}\left| n\right\rangle,\quad\text{where}\,\,\alpha\in\mathbb{C}\, .
\end{equation}
They overlap as 
\begin{equation}
\label{overlapCS1}
\lg\alpha|\alpha^{\prime}\rg:=e^{-\frac{\vert\alpha\vert^{2}}{2}-\frac{\vert\alpha^{\prime}\vert^{2}}{2}+\bar\alpha\alpha^{\prime}}= e^{\ii \mathrm{Im}\left(\bar\alpha\alpha^{\prime}\right)}e^{-\frac{\vert\alpha -\alpha^{\prime}\vert^2}{2}}\, .
\end{equation}
For an alphabet of two GS-CS  given by
\begin{equation}
\label{alphLCS}
\rho=|\alpha\rangle\langle \alpha|\, ,\quad
\rho^{\prime}=|\alpha^{\prime}\rangle\langle \alpha^{\prime}|\, ,
\end{equation}
we have from \eqref{overlapCS1}  
\begin{equation}
\label{over1}
\mathrm{tr}\left(\rho\rho^{\prime}\right)= \left|\langle\alpha|\alpha^{\prime}\rangle \right|^2=e^{-\vert \alpha-\alpha^{\prime}\vert^2}= \vert\lg\alpha-\alpha^{\prime}|0\rg\vert^2\, .
\end{equation} 
As was pointed  out in the introduction, this property of the  GS-CS allows to restrict our study to an alphabet of two GS-CS  given by 
\begin{equation}
\label{alphLCS}
\rho_0=|0\rangle\langle 0|\, ,\quad
\rho_1=|\alpha\rangle\langle \alpha|\, ,
\end{equation}
for which 
\begin{equation}
\label{over1}
\left|\langle\alpha|0\rangle \right|^2=e^{-|\alpha|^2}\, .
\end{equation}
Now the quantity $u=|\alpha|^2$ is precisely the average value of the number operator $\hat N=\sum_{n=0}^{\infty} n|n\rg\lg n|$, i.e. the mean value of number of photons in the CS $|\alpha\rangle$,
\begin{equation}
\label{meannCS}
\bar n = \lg \alpha | \hat N |\alpha \rg = \sum_{n=0}^{\infty} n \frac{u^n e^{-u}}{n!} =u\,.
\end{equation} 
Note that $u$ can also be viewed as the mean value of the random variable $n$ with the Poissonian  probability distribution
\begin{equation}
\label{poissondist}
\bar n = \sum_{n=0}^{\infty} n \sfP_n(u)\, \qquad \sfP_n(u) = \left|\langle\alpha|n\rangle \right|^2 = e^{-u}\frac{u^n }{n!} .
\end{equation}
Hence \eqref{over1} becomes 
\begin{equation}
\label{over2}
\left|\langle\alpha|0\rangle \right|^2=e^{-\bar n}\, ,
\end{equation}
and the quantum error probability reads in this case
\begin{equation} \label{helstromGSCS}
  P_\mathrm{H}  = \frac{1}{2} \left(1-\sqrt{1-4 \xi_0 \xi_1 e^{-\bar n}  }\right) \equiv P_\mathrm{H} (\bar n) \,.
\end{equation}

\subsubsection{Imperfect detection}

  In practice, only a fraction $\eta$ of the photons reaching a photo-counter leads to a \textit{count}. The parameter $\eta\in[0,1]$ is called the efficiency parameter.
For imperfect detection ($\eta<1$), the binomial distribution allows to compute the probability  $\sfP_n(u; \eta)$ to detect $n$-photons as a function of $\eta$. {Sub-unity photodetector efficiency yields
a photon-count which is linked to the perfect ($\eta = 1$) photon distribution} $\sfP_m(u)=\sfP_m(u;\eta = 1)$ 
through a Bernoulli transformation \cite{main:loudon}: 
 \begin{equation}
\label{probnideal}
\sfP_n(u;\eta) = \sum_{m=n}^{\infty}\binom{m}{n} {\eta^n}\, (1-\eta)^{m-n}\,\sfP_m(u;\eta = 1).
\end{equation}
The GS-CS's { exhibit the useful property that non-unit quantum efficiency corresponds to a perfect
detector equipped with a beam-splitter having a transmission coefficient $\eta \le 1$. In fact, since one has the Poisson distribution for GS coherent states and an ideal detector, \eqref{poissondist}, replacing $m$ by $s = m-n$ in the summation (\ref{probnideal}) leads to}
\begin{equation}
\label{probnidealcs}
\sfP_n(u;\eta) = \frac{\eta^n u^n}{n!} \, e^{-\eta u}= \frac{(\eta u)^n }{n!} \, e^{-\eta u} = \sfP_n(\eta u)\,.
\end{equation}
In other words, one can ``hide'' the imperfection of the detector into an effective state
by changing
\begin{equation}
\label{etaalpha}
\alpha \to \sqrt{\eta}\alpha.
\end{equation}
This amounts to replace the previous setup based on the alphabet $\{|0\rangle, |\alpha\rangle\}$ and an imperfect detector ($\eta<1$) with a new one based on $\{|0\rangle , |\sqrt{\eta}\alpha\rangle\}$ and a perfect detector \cite{geremia04}. Then, the overlap of the new alphabet reads
\begin{equation}
\label{over3}
\left| \langle \sqrt{\eta}\alpha|0\rangle \right|^2=e^{-\eta \bar n}\, ,
\end{equation}
where $\eta \bar n$ is equal to the mean photon number given by the modified probability distribution \eqref{probnidealcs}.
\begin{equation}
\label{newnbar}
\bar n_{\eta} = \sum_{n=0}^{\infty} n \, \sfP_n(u;\eta)= \eta u = \eta \bar n\,.
\end{equation}
This allows to write down the Helstrom bound \eqref{Equation::PerfectHelstrom} for the imperfect case as
\begin{equation} \label{Equation::RealHelstrom}
  P_\mathrm{H}(\bar n;\eta) 
   = \frac{1}{2} \left(1-\sqrt{1-4 \xi_0 \xi_1 e^{-\bar{n}_\eta} } \right) \,.
\end{equation}
This indicates that there is always a non-zero quantum error probability with GS-CS.

\section{A class of non standard coherent states in Quantum Optics: the AN-CS}
\label{ANclass}
Most of the generalisations of GS-CS  encountered in quantum optics or quantum mechanics are  coherent states in the AN class (AN for ``$\alpha n$'') \cite{gazeau19}. These ``AN-CS'' are one-mode Fock states  of the  form;
\begin{equation}
\label{anclass}
|\alpha;\bsh\rg = \sum_{n=0}^{\infty} \alpha^n\,h_n(\vert\alpha\vert^2) |n\rg\, , 
\end{equation}
where the complex $\alpha$ lies in the open disk $\mathcal{D}_R$ of radius $R$, $\vert \alpha\vert<  R$, with $R$ being finite or infinite. The sequence $\bsh:= \left( h_n\right)\, , \, n=0,1,2,\dotsc$ of real-valued functions 
\begin{equation}
\label{hnu}
[0,R^2) \ni u :=\vert \alpha\vert^2 \mapsto h_n(u)
\end{equation}
is requested to obey the three fundamental conditions:
\begin{align}
\label{condnorm}
    &    \sum_{n=0}^{\infty} u^n\,\left(h_n(u)\right)^2\, =1 , \quad \mbox{(normalisation 1)}\, ,  \\
 \label{onetoone} [0,R^2) \ni u \mapsto \bar n_{\bsh} (u)& := \sum_{n=0}^{\infty} n\, u^n\,\left(h_n(u)\right)^2 \quad \mbox{is strictly increasing}\, ,\\
\label{condortho} \exists w: [0,R^2]\to {\mathbb R}^+  \ \mbox{such that}\     &  \int_{0}^{R^2}\ud u\,w(u)\,u^n\left( h_n(u)\right)^2 \, =1, 
\quad \mbox{(normalisation 2)}\, ,
\end{align}
where $w$ is a weight function. 

Note that the GS-CS's are the particular case
\begin{equation}
\label{gscspc}
h_n(u)= \frac{e^{-u/2}}{\sqrt{n!}}\, , \quad w(u)=1\, \,  ,\,\, \,\,\,\, R = \infty \, . 
\end{equation}
{A finite summation in (3.1) due to $h_n = 0$ for n larger than some 
$n_{\mathrm{max}}$ might be considered in the present study.} 
In this case we consider 
as Hilbert space the finite dimensional Hilbert space spanned by the 
orthonormal basis ($\{ \vert n \rangle \} )_{n=0}^{n_{\mathrm{max}}}$. Spin CS's in Subsection \ref{spinCS} are an illustration of this case. 

In the infinite sum case, special conditions are to be imposed on the sequence of functions $h_n$ in order to insure the convergence of series \eqref{condnorm} for all $0\leq u <R^2$ and integrals \eqref{condortho} for all $n$.

Condition  Eq. \eqref{condnorm} means that the vectors \eqref{anclass} are normalised states {of a Fock Hilbert space formed by the number states. This yields a probability distribution on $\N$
\begin{equation}
\label{probdet}
n \mapsto   \sfP^{\bsh}_n(u)  = \left|\langle\alpha; \bsh|n\rangle \right|^2 = u^n\,\left(h_n\left(u\right)\right)^2
\end{equation}
where $u$ is now a parameter. This is exactly the probability} of registering $n$ photons with a measuring ideal device, i.e., having maximal efficiency ($\eta=1$) when the  light beam is in the non standard  coherent state $|\alpha;\bsh\rg$.

Condition  Eq. \eqref{onetoone} expresses that  the expected value 
\begin{equation}
\label{avernb}
\lg\alpha;\bsh| \hat N |\alpha;\bsh\rg = \sum_{n=0}^{+\infty} n \sfP^{\bsh}_n(u)\equiv \bar n_{\bsh} (u) 
\end{equation}
of the number operator in AN-CS is a one-to-one relation with $u$ and so can be inverted. 

Condition  Eq. \eqref{condortho} implies the resolution of the identity in the same Fock space:
\begin{equation}
\label{resid}
 \int_{\mathcal{D}_R}\frac{\ud^2\alpha}{\pi}\,w\left(\vert\alpha\vert^2\right)\, |\alpha;\bsh\rg\lg\alpha;\bsh| =\id\, .
\end{equation}
This property holds because of the orthonormality relations 
\begin{equation}
\label{orthophi}
 \int_{\mathcal{D}_R}\frac{\ud^2\alpha}{\pi}\,w\left(\vert\alpha\vert^2\right)\,\bar{\alpha}^{n^{\prime}}\alpha^n h_{n^{\prime}}\left(\vert\alpha\vert^2\right)\,h_n\left(\vert\alpha\vert^2\right)= \delta_{n n^{\prime}}
\end{equation}
stemming from Fourier angular integration on the argument of $\alpha$ and  the  kind of moment problem solved by \eqref{condortho}.
The latter  allows us to interpret the map 
\begin{equation}
\label{alprob}
\alpha \mapsto  \sfP^{\bsh}_n\left(\vert\alpha\vert^2\right)
\end{equation}
as an isotropic  probability distribution, with parameter $n$,  on the disk $\mathcal{D}_R$, equipped with the  measure $w\left(\vert\alpha\vert^2\right)\,\dfrac{\ud^2\alpha}{\pi}$. Equivalently, the map
\begin{equation}
\label{uprob}
[0,R^2) \ni u \mapsto   \sfP^{\bsh}_n(u) 
\end{equation}
is a probability distribution on the interval $[0,R^2)$ equipped with the measure $w(u)\ud u$.

 The average value \eqref{avernb} of the number operator 
{ might be interpreted as the intensity (or energy if multiplied by $\hbar \omega$)  of the  quantum radiation (monochromatic) in the state $|\alpha;\bsh\rg$. Thus, an optical phase space related to this ``AN-CS radiation'' might be constructed through  the map }
 \begin{equation}
\label{mapaln}
\mathcal{D}_R \ni \alpha \mapsto \zeta_{\alpha}=\sqrt{\bar n_{\bsh}\left(\vert\alpha\vert^2\right)}\,e^{\ii \arg \alpha}\in \C\, .
\end{equation}

The fluctuations of the number of photons  about its mean value 
are quantified in terms of the standard deviation 
\begin{equation}
\Delta n_{\bsh}(u)=\sqrt{\overline{n^2_{\bsh}}-\left(\bar{n}_{\bsh} \right)^2}\, , \quad \overline{n^2_{\bsh}}(u)=\sum_n n^2 \sfP^{\bsh}_n(u)\,.
\end{equation}

The distribution $n \mapsto  \sfP^{\bsh}_n(u)$ is then classified as: sub-Poissonian for $\Delta n_{\bsh}<\sqrt{\bar{n}_{\bsh}}$, Poissonian for $\Delta n_{\bsh}=\sqrt{\bar{n}_{\bsh}}$ and super-Poissonian for $\Delta n_{\bsh}>\sqrt{\bar{n}_{\bsh}}$.
The deviation of $\sfP^{\bsh}_n(u)$ from the Poisson distribution, is conveniently measured with the \textit{Mandel parameter} $Q^{\bsh}_M$. The latter is defined as
\begin{equation}\label{mandelQu}
\begin{aligned}
Q^{\bsh}_M(u):=\frac{\overline{n^2_{\bsh}}-\left(\bar{n}_{\bsh} \right)^2}{\bar{n}_{\bsh}}-1\, , 
\end{aligned}
\end{equation}
and it is clear to see that it should be larger than or equal $-1$. 
At a given $u$  the distribution $\sfP^{\bsh}_n(u)$ is respectively Poissonian if $Q^{\bsh}_M(u)=0$, sub-Poissonian if $Q^{\bsh}_M(u)<0$, and super-Poissonian if $Q^{\bsh}_M(u)>0$. Actually, due to the bijective relation between $u$ and $\bar n$ according to Assumption \eqref{onetoone},  we will preferably analyse the behaviour of the Mandel parameter in function of the physical $\bar n$:
\begin{equation}
\label{mandelQn}
\widetilde{Q^{\bsh}_M}(\bar n)= Q^{\bsh}_M(u(\bar n))\,.
\end{equation}

\section{Helstrom bound for states in the AN-class versus GS-CS}
\label{HB-AN}
\subsection{Perfect detection}
The Helstrom bound (HB) for an alphabet composed of pure states depends only on the overlap between them, as it is obvious from \eqref{Equation::PerfectHelstrom}. If one wishes to lower the HB in comparison with the HB \ref{helstromGSCS} involving GS-CS, it is necessary to lower  the overlap of the two involved states. For this reason we are interested in states with a non-Poissonian photon distribution. Since our objective  is to compare their corresponding HB with \eqref{helstromGSCS} and \eqref{Equation::RealHelstrom}, we restrict our study to the overlap between the vacuum and an arbitrary AN-CS.
Accordingly, for an alphabet of  two AN-CS's  given respectively by
\begin{equation}
\label{alphNLCS}
\rho_0=|0;\bsh\rangle\langle 0;\bsh|\, ,\quad 
\rho_1=|\alpha;\bsh\rangle\langle\alpha; \bsh|\, ,
\end{equation}
the modulus squared of the overlap reads
\begin{equation}
\label{over2}
\left|\langle\alpha;\bsh|0;\bsh\rangle \right|^2=\left( h_0( u)\right)^2\, .
\end{equation}
Then the corresponding HB  reads, 
\begin{equation}
\label{HBAN}
P^{\bsh}_\mathrm{H} = \frac{1}{2} \left(1-\sqrt{1-4 \xi_0 \xi_1  
  \left( h_0( u)\right)^2} \right).
\end{equation}
Comparing two Helstrom bounds is physically meaningful  only if one views them as a function of the average number of photons $\overline{n}$. This is made possible thanks to Condition \eqref{onetoone} which allows us to express unambiguously the inverse function $\bar n \mapsto u^{\bsh}(\bar n)$ from Eq. \eqref{avernb}. Hence, $P^{\bsh}_\mathrm{H}$ in   Eq. \eqref{HBAN} can be written as a function of $\bar n$
\begin{equation}
\label{HBANn}
\bar n \mapsto P^{\bsh}_\mathrm{H}(\bar n) = \frac{1}{2} \left(1-\sqrt{1-4 \xi_0 \xi_1  
  \left( h_0\left( u^{\bsh}(\bar n)\right)\right)^2} \right)\,.
\end{equation}
Comparing this HB with the standard one given by \eqref{helstromGSCS} amounts to study the function
\begin{equation}
\label{Deltah}
\Delta^{\bsh}(\bar n) :=  \left( h_0\left( u^{\bsh}(\bar n)\right)\right)^2- e^{-\bar n} \,.
\end{equation}
Indeed, we have 
\begin{equation}
\label{PhHPh}
\Delta^{\bsh}(\bar n) \lesseqqgtr 0 \Longleftrightarrow P^{\bsh}_\mathrm{H}(\bar n)  \lesseqqgtr  P_\mathrm{H}(\bar n)\,.
\end{equation}

\subsection{Imperfect detection}

{In the case of imperfect detection,  the probability
$\sfP^{\bsh}_n(u;\eta)$ to detect $n$-photons using a non-perfect detector
($\eta < 1$) is given in terms of the (perfect detector) probability,
$\sfP^{\bsh}_m(u)=\sfP^{\bsh}_m(u;\eta = 1),$ } through the
Bernoulli transformation \eqref{probnideal}
 \begin{equation}
\label{probnidealh}
\sfP^{\bsh}_n(u;\eta) = \sum_{m=n}^{\infty}\binom{m}{n} \eta^n\, (1-\eta)^{m-n}\,\sfP^{\bsh}_m(u;\eta = 1)
\end{equation}
This transformation has the following important property. The mean value of a function $n\mapsto \varphi(n)$ with respect to the above distribution is  given by:
\begin{equation}
\label{avervarphi}
\overline{\varphi(n)}(u;\eta)= \sum_{n=0}^{+\infty}\varphi(n)\,\sfP^{\bsh}_n(u;\eta)= \sum_{n=0}^{+\infty}\lg\varphi\rg_{n;\eta}\,\sfP^{\bsh}_n(u;\eta=1)\, , 
\end{equation}
where $\lg\cdot\rg_{n;\eta}$ is the mean value with respect to the binomial distribution with parameter $\eta$ and for $n$ trials,
\begin{equation}
\label{meanbin}
\lg\varphi\rg_{n;\eta} =  \sum_{k=0}^{n}\binom{n}{k} \eta^k\, (1-\eta)^{n-k}\, \varphi(k) \, .
\end{equation}
For the number function, $\varphi(k)=k$, $\lg\varphi\rg_{n;\eta} =\eta n$, and we end up with the simple expression
\begin{equation}
\label{averneeta}
\bar n(u;\eta)= \eta\bar n(u;\eta=1) \equiv \eta\bar n(u)\,. 
\end{equation}
There results that the only change we have to do in the expression \eqref{Deltah} and the inequalities \eqref{PhHPh} is to replace $\bar n$ with $\eta\bar n$:
\begin{equation}
\label{Deltaheta}
\Delta^{\bsh}(\eta\bar n) :=  \left( h_0\left( u^{\bsh}(\eta\bar n)\right)\right)^2- e^{-\eta\bar n} \,,
\end{equation} 
and
\begin{equation}
\label{PhHPheta}
\Delta^{\bsh}(\eta\bar n) \lesseqqgtr 0 \Longleftrightarrow P^{\bsh}_\mathrm{H}(\eta\bar n)  \lesseqqgtr  P_\mathrm{H}(\eta\bar n)\,.
\end{equation}
It is clear from the above that the behavior of the Helstrom bound with regard to the GS-CS case is not $\eta$-dependent.
Note that the relation \eqref{averneeta} reduces to the rescaled $\alpha$ parameter  \eqref{over3} in the case of the standard GS-CS. 

\section{Helstrom bound with ``non-linear'' coherent states}
\label{HBNLCS}
{We define non-linear CS (NL-CS) as AN-CS for with functions $h_n(u)$ assuming the deformed Poissonian form}
\begin{equation}
\label{nlgpn}
h_n(u)= \frac{\lambda_n}{\sqrt{\mathcal{N}(u)}}\equiv \frac{1}{\sqrt{\mathcal{N}(u)}}\,\frac{1}{\sqrt{x_n!}} \, , \quad x_0!:=1\, , \quad x_n! := x_1x_2\cdots x_n\, , 
\end{equation}
where the $\lambda_n$'s  form a  strictly decreasing sequence of positive numbers,
\begin{equation}
\label{defpoiss}
\boldsymbol{\lambda}:=(\lambda_n)_{n\in \N}\,, \quad \lambda_0 = 1 \,,   
\end{equation}
i.e.,  the $x_n$'s, $x_n:=  \left(\frac{\lambda_{n-1}}{\lambda_n}\right)^2$,  form the  strictly increasing sequence
\begin{equation}
\label{xnxn1}
\boldsymbol{\chi}:=(x_n)_{n\in \N}\,, \quad 0= x_0 <x_1<x_2 \cdots x_n <x_{n+1}< \cdots\, . 
\end{equation}
{The function $\mathcal{N}(u)$ is the generalized exponential with radius $R^2$,}
\begin{equation}
\label{genexp}
\mathcal{N}(u) = \sum_{n}  \lambda_n^2 u^n = \sum_{n} \frac{u^n}{x_n!}\, ,
\end{equation}
The corresponding NL-CS read
\begin{equation}
\label{nlcs}
|\alpha;\boldsymbol{\lambda}\rg= |\alpha;\boldsymbol{\chi}\rg= \frac{1}{\sqrt{\mathcal{N}\left(\vert\alpha\vert^2\right)}}\sum_{n} \lambda_n \alpha^n|n\rg=\frac{1}{\sqrt{\mathcal{N}\left(\vert\alpha\vert^2\right)}}\sum_{n}\frac{\alpha^n}{\sqrt{x_n!}}|n\rg\,.
\end{equation}
Here, for the sake of simplicity, we drop subscript $\bsh$ or $\boldsymbol{\lambda}$ or $\boldsymbol{\chi}$.
The normalisation  condition \eqref{condortho} is fulfilled if there exists a weight $w(u)$ solving   the  Stieltjes moment problem \cite{akhiezer65} for the sequence $(x_n!)_{n\in \N}$, 
\begin{equation}
\label{mompbxn}
x_n!= \int_0^{R^2}\ud u\,\frac{w(u)}{\mathcal{N}(u)}\, u^n\,. 
\end{equation}
{The detection probability follows the deformed Poisson distribution: }
\begin{equation}
\label{poissondef}
n\mapsto \sfP_n(u)  = \left|\langle\alpha; \boldsymbol{\chi} |n\rangle \right|^2
= \frac{1}{\mathcal{N}(u)}\,\frac{u^n}{x_n!}\,. 
\end{equation}
The average value $\bar n$ of the number operator, as a function of $u=\vert \alpha \vert^2$, is given by
\begin{equation}
\label{nlcsavern}
\bar n (u)= \lg \alpha|\hN|\alpha\rg= u \frac{\mathcal{N}^{\prime}(u)}{\mathcal{N}(u)}= u\frac{\ud \log\mathcal{N}(u)}{\ud u}\,. 
\end{equation}
The above relation  imposes that 
\begin{equation}
\label{Nppos}
\mathcal{N}^{\prime}(u)>0\,,
\end{equation}
i.e., $\mathcal{N}(u)$ is strictly increasing from $\mathcal{N}(0)=1$ over $\R_+$. 

Now, Condition \eqref{onetoone} imposes a further relation on nonlinear CS:
\begin{equation}
\label{onetoonenl}
\frac{\ud \bar n}{\ud u} = \left(\frac{d}{du} + u\, \frac{d^2}{du^2}\right) \ln \mathcal{N}(u) \, > 0 .
\end{equation}
The expected value $\overline{n^2}$ is given by
\begin{equation}
\label{nclsavern2}
\overline{n^2}(u)=\frac{u}{\mathcal{N}(u)}\frac{\ud}{\ud u}u\frac{\ud}{\ud u}\mathcal{N}(u)\, ,
\end{equation}
and the Mandel parameter $Q_M(u)$ reads
\begin{equation}
Q_M(u)=u\left(\frac{\mathcal{N}^{\prime\prime}(u)}{\mathcal{N}^{\prime}(u)}-\frac{\mathcal{N}^{\prime}(u)}{\mathcal{N}(u)}\right)=u\frac{\ud}{\ud u}\ln\left(\frac{\ud}{\ud u}\ln \mathcal{N} (u)\right)\,.
\end{equation}
Let us put
\begin{equation}
\label{NuFu}
\mathcal{N}(u)= e^{F(u)}\,, \quad F(0) =0\,.
\end{equation}
Since for $u>0$,  $0< \mathcal{N}^{\prime}(u) = F^{\prime}(u)e^{F(u)}$, we have $F^{\prime}(u) >0$ for $u>0$, i.e. $F(u)$ is strictly increasing over $\R_+$ from $F(0) =0$. Note that $F^{\prime}(u) = \dfrac{\bar n(u)}{u}\,$.
In term of $F$, the Mandel parameter assumes the simpler form:
\begin{equation}
\label{QMF}
Q_M(u)= u\,\frac{F^{\prime\prime}(u)}{F^{\prime}(u)}= u\frac{\ud }{\ud u}\ln F^{\prime}(u)\,.
\end{equation}
In this case, \eqref{onetoonenl} leads to $u\,\dfrac{F^{\prime\prime}}{F^{\prime}}= u\, \dfrac{\ud \ln F^{\prime}}{\ud u} > -1\,$, which is consistent with the known lower bound on Mandel parameter.
Hence, the NL-CS \eqref{nlcs} are Poissonian if and only if $F^{\prime\prime}(u)=0$ for all $u\in \R_+$, i.e. if they are the GS-CS. They are sub-Poissonian (resp. super-Poissonian) at $u$ if $F^{\prime\prime}(u)<0$ (resp. $>0$). 

Finally, applying the expression \eqref{Deltah} to compare the Helstrom Bounds between the above NL-CS \eqref{nlcs} and the GS-CS leads to
\begin{equation}
\label{Deltahnl}
P^{\chi}_\mathrm{H}(\bar n)  \lesseqqgtr  P_\mathrm{H}(\bar n)\Longleftrightarrow \Delta(\bar n) =  e^{-F(u(\bar n))}- e^{-\bar n} \lesseqqgtr 0 \Longleftrightarrow  \bar n \lesseqqgtr F(u(\bar n))\,.
\end{equation}

\section{Mandel Parameter and Helstrom Bound for selected NL-CS families}
\label{manhelnlex}
\subsection{Spin CS as optical CS}
\label{spinCS}
\begin{figure}[H]
	\centering
	\includegraphics[width=0.45\textwidth]{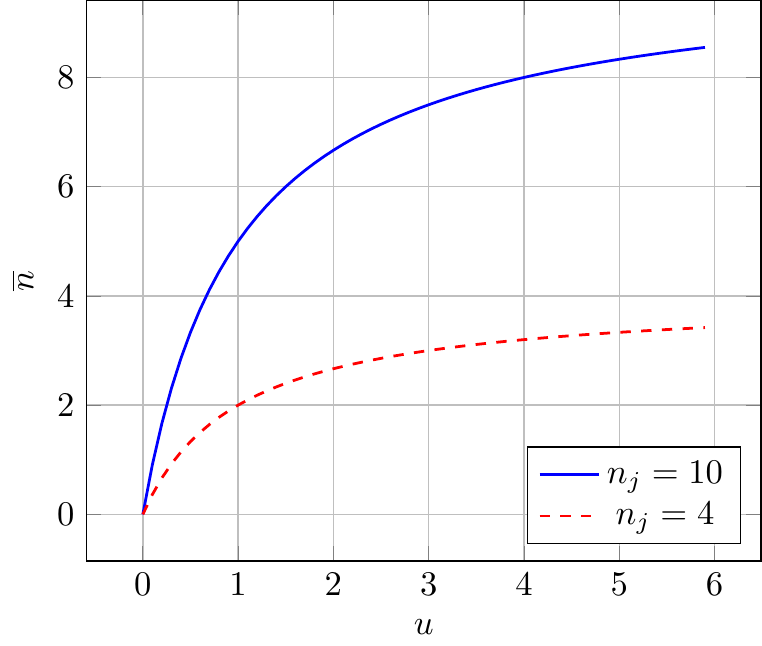}
	\includegraphics[width=0.45\textwidth]{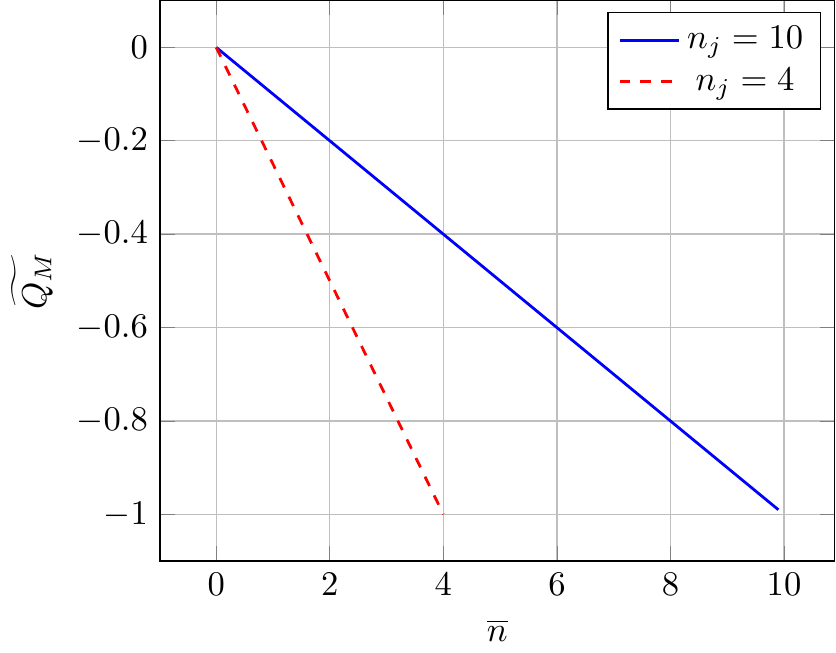}
	\caption{Photon statistics for optical spin CS  \eqref{zetaspincs} with $n_j= 4$ and $10$. These states  are always sub-Poissonian. Left figure: Average value of photons as a function of $u$. Right figure: Mandel parameter $\widetilde{Q_M}$ \eqref{spinMand} as a function of $\bar n$.   }
	\label{fig:Spin_Mandel}
\end{figure}

{This set of states derives from the Gilmore or Perelomov SU$(2)$-CS, also called spin CS \cite{perel72,perel86},  adapted to the context of quantum optics.} The Fock  space reduces to the finite-dimensional  subspace $\mathcal{H}_j$, with dimension $n_j+1:=2j+1$, { $j$ being a positive integer or half-integer.} 
They are defined through \eqref{anclass} and \eqref{nlgpn} with
\begin{equation}
\label{binomialcase}
h_n(u)= \lambda_n\, (1 + u)^{-\frac{n_j}{2}}\,, \quad \lambda_n=\sqrt{\binom{n_j}{n}}\,, \quad \binom{n_j}{n}=\frac{n_j!}{n!(n_j-n)!}\, ,\quad  x_n= \frac{n}{n_j-n+1}\,, 
\end{equation}
for $n\in[0,n_j]$; $h_{n> n_j}(u)=0$. 
The corresponding generalised exponential  is the binomial 
\begin{equation}
\label{spinbin}
\mathcal{N}(u)= (1 + u)^{n_j}\,,
\end{equation}
and the related NL-CS read as
\begin{equation}
\label{zetaspincs}
|\alpha;n_j\rg =  \left(1 + \vert \alpha \vert^2\right)^{-\frac{n_j}{2}}\sum_{n=0}^{n_j} \sqrt{\binom{n_j}{n}}\, \alpha^{n}\, |n\rg\,.
\end{equation}
They resolve the unity in $\mathcal{H}_{n_j}$ in the following way:
\begin{equation}
\label{csresunzeta}
\frac{n_j+1}{\pi}\, \int_{\C}\frac{\ud^2\alpha}{(1 + \vert \alpha\vert^2)^2}\, |\alpha;n_j\rg \lg \alpha;n_j|  =\id\,.
\end{equation}
{The  detection probability follows the  binomial distribution: }
\begin{equation}
\label{binomialdist}
n\mapsto \sfP_n(u)= (1+u)^{-n_j}\,\binom{n_j}{n}\,u^n\,. 
\end{equation}
{The number operator average value then reads}
\begin{equation}
\label{avernphspin}
\bar n (u)= n_j\,\frac{u}{1+u}\ \Leftrightarrow\ u= \frac{\bar n/n_j}{1-\bar n/n_j} \,,
\end{equation}
which is depicted in Figure  \ref{fig:Spin_Mandel} for $n_j=4$ and $ 10$.
The probability \eqref{binomialdist} is thus expressed in terms of 
$p:=\bar n/n_j$ as 
 \begin{equation}
\label{binomialdist1}
\sfP_n(u) \equiv \widetilde{\sfP}_n(p)= \binom{n_j}{n}\,(1-p)^{n_j-n} \, p^n\, .
\end{equation}
{As a consequence one can  build an optical phase space as the open disk of radius $\sqrt{n_j}$, }
$$\mathcal{D}_{\sqrt{n_j}}= \left\{\zeta_{\alpha}= \sqrt{\bar n\left(\vert \alpha\vert^2\right)}e^{\ii \arg \alpha}\, ,\,  \vert \zeta_{\alpha}\vert < \sqrt{n_j}\right\}\,.$$

{
In terms of photon statistics, the physical interpretation of the binomial distribution together with $n_j$  is decidedly clear when considering a  beam of light which is perfectly coherent and has constant intensity. For a beam of a finite length $L$ which is divided in $n_j$ parts of length $L/n_j$, $\widetilde{\sfP}_n(p)$ stands for  the probability to find, in any possible order, one photon in $n$ segments and no photon in the remaining $(n_j - n)$ segments \cite{fox06}.   
A more general  interpretation of the above distributions \eqref{binomialdist} and \eqref{binomialdist1} can be found in \cite{algahel08}. }

Note that the GS-CS {derive from the above states in the limit $n_j \to \infty$. This goes throughout a contraction process which consists in rescaling the complex variable $\alpha$ through
$\alpha \mapsto\sqrt{n_j}\, \alpha$. As a consequence the distribution $\widetilde{\sfP}_n(p)$ reduces to \eqref{poissondist}. 
%
%
}

By applying \eqref{QMF} with $F(u) = n_j\ln(1+u)$, the expression of the Mandel parameter (as a function of $u$ or as a function of $\bar n$) is independent of $j$ and is given  by
\begin{equation}
\label{spinMand}
Q_M(u)= -\frac{u}{1+u}\ \Leftrightarrow\ \widetilde{Q}_M(\bar n)= -\frac{\bar n}{n_j}\,.
\end{equation}
Hence,  states \eqref{zetaspincs} are always sub-Poissonian, as shown in Figure \ref{fig:Spin_Mandel}. This fundamentally quantum feature is easily understood since they are built from a finite number of photons.\\

\begin{figure}[H]
	\centering
	\includegraphics[width=0.45\textwidth]{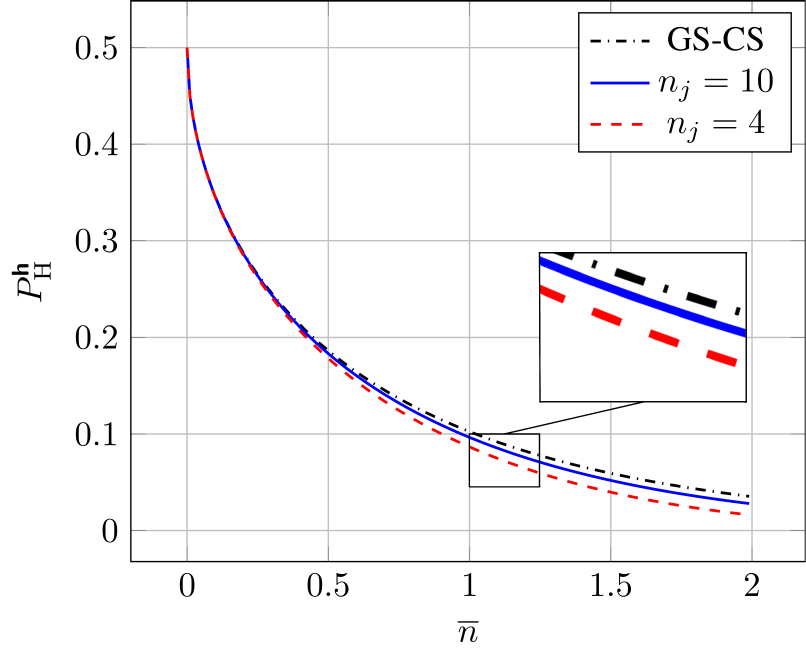}
	\includegraphics[width=0.45\textwidth]{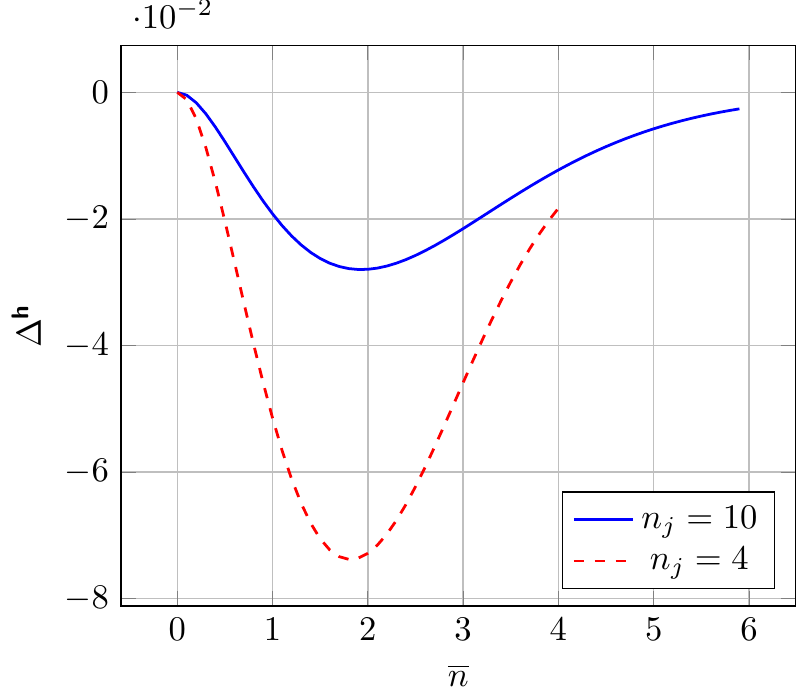}
	\caption{Helstrom bounds are plotted for GS-CS and spin CS  with $n_j= 4$ and $10$. The spin SC Helstrom bound is slightly lower,  for any value $\bar n\leq n_j$. Left: Helstrom bound as function of $\bar{n}$. Right: Delta function,  (\ref{persu11helst}) versus $\bar{n}$.}
	\label{fig_SHB_spin}
\end{figure}

Concerning the Helstrom bound, we examine the expression \eqref{Deltah} in the present case:
\begin{equation}
\label{spinhelst}
\Delta(\bar n)= \left(1-\frac{\bar n}{n_j}\right)^{n_j} - e^{-\bar n}\,.
\end{equation}
 From the Bernouilli-like inequality 
 \begin{equation}
\label{berineq}
(1+x)^r\leq e^{rx}\, , 
\end{equation}
valid for any $r>0$ and any real $x$, we infer that $\Delta(\bar n)\leq 0$ for any $\bar n$ and $j$. Hence the spin CS Helstrom bound is lower than the GS-CS Helstrom bound  for any value of $u=\vert \alpha\vert^2$, see  Figure \ref{fig_SHB_spin}. This is a remarkable result which confirms their deep quantum nature.

\subsection{SU$(1,1)$-CS as optical CS}
\label{SUIICS}
The application in quantum optics of coherent states built from representations of the non-compact group SU$(1,1)$  lying in the discrete series has been carried out  by various authors, {see for instance the noticeable papers by Wodkiewicz and Eberly \cite{wodkiewicz85}, Gerry \cite{gerry87}, Brif \cite{brif95}. See also in \cite{hachetal16,hachetal18} (and references therein) the interesting properties revealed  by  such states when they are entangled.}

\subsubsection{Perelomov CS} \label{persu11}
 {This set of states derives from  Perelomov SU$(1,1)$-CS \cite{perel72,perel86}, here again adapted to the context of quantum optics. They emerge through the unitary action of SU$(1,1)$ on number states. This leads to an infinite-dimensional Fock Hilbert space $\mathcal{H}$ with $\alpha$  constrained to the open disk, $\mathcal{D}:= \{\alpha  \in \C\, , \, \vert \alpha \vert < 1\}$. 
}
When the Fock state is the vacuum and  $\varkappa> 1/2$,  the Perelomov $\varkappa$-dependent CS are the following NL-CS:
\begin{equation}
\label{cssu11}
 |\alpha;\varkappa\rg_{\mathrm{Per}}= \sum_{n=0}^{\infty}\alpha^n\,h_{n}\left(\vert\alpha\vert^2\right)\,|n\rg\, , \quad h_{n}(u)=  \lambda_n (1-u)^{\varkappa}\,, \quad \lambda_n= \sqrt{\binom{2\varkappa -1 +n}{n}}\, ,\quad x_n= \frac{n}{2\varkappa-1+n}\,. 
\end{equation}
The corresponding generalised exponential  is the binomial (also called $q$-exponential in different contexts, {see \cite{bercugaro12} and references therein):}
\begin{equation}
\label{su11bin}
\mathcal{N}(u)= (1 - u)^{-2\varkappa}\,.
\end{equation}
They  solve the identity:
\begin{equation}
\label{runitmorecssu11}
\frac{2\varkappa-1}{\pi}\int_{\mathcal{D}}\frac{\ud^2 \alpha}{\left(1-\vert\alpha\vert^2\right)^2} \, |\alpha; \varkappa\rg_{\mathrm{Per}} {}_{\mathrm{Per}}\lg \alpha; \varkappa | = \id\,.
\end{equation}

{This results in a negative binomial distribution, \cite{algahel08},}
\begin{equation}
\label{negbinomialdist}
n\mapsto \sfP_n(u)= (1-u)^{2\varkappa}\,\binom{2\varkappa -1 +n}{n}\,u^n\,. 
\end{equation}
{While the number operator average value takes the form}
\begin{equation}
\label{avernphperel}
\bar n(u)= 2\varkappa\,\frac{u}{1-u}\ \Leftrightarrow\ u= {\frac{\bar  n/2\vk}{1+\bar n/2\vk}} \,. 
\end{equation}
Note that introducing the efficiency {$\eta := \dfrac{1}{2\vk}\in (0,1)$} allows to express the probability \eqref{negbinomialdist} { as a function of the scaled average photocount number $\bar N:=\eta\bar n$, }
 \begin{equation}
\label{negbinomialdist1}
\sfP_n(u) \equiv \widetilde{\sfP}_n(\bar N)= (1+\bar N)^{-1/\eta}\binom{1/\eta-1 +n}{n}\, \left(\frac{\bar N}{1+\bar N}\right)^n\, .
\end{equation}
{This distribution has a notable property to reduce to Bose-Einstein distribution at the bound $\eta= 1$, which corresponds to the limit $\vk = 1/2$  of SU$(1,1)$ discrete series. 
At sub-unit efficiency, $\eta< 1$, i.e. $\vk>1/2$, deviation from the Bose-Einstein distribution can be interpreted as resulting from using the scaled average value $\bar N$ (detected photons) in place of the mean value $\bar n$ of photons effectively reaching the detector (not all detected) \cite{fox06, aharonov73}.
}

\begin{figure}[htb]
	\centering
	\includegraphics[width=0.45\textwidth]{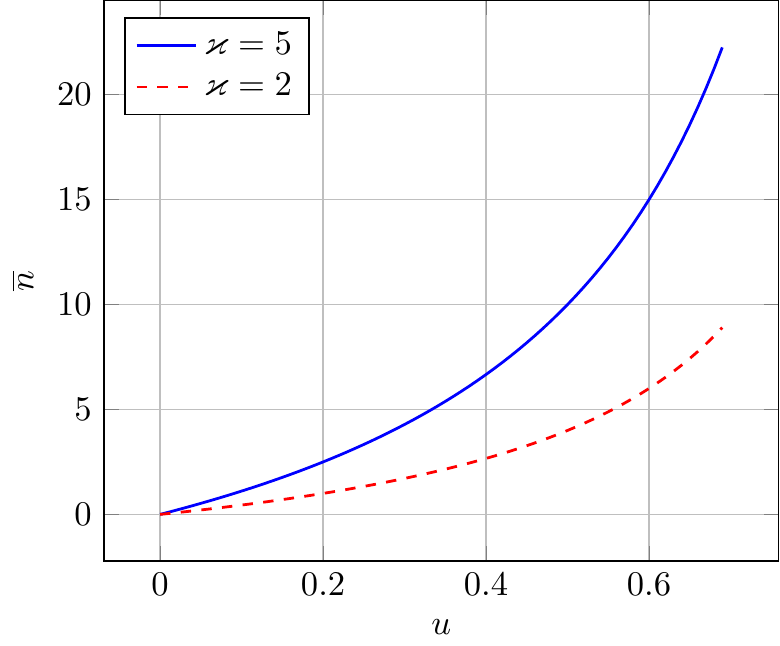}
	\includegraphics[width=0.45\textwidth]{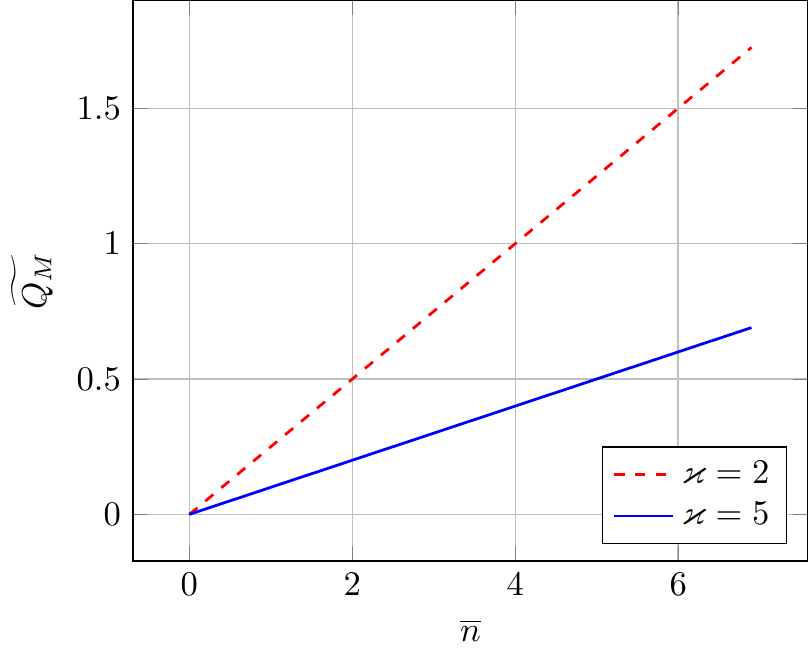}
	\caption{Photon statistics for Perelomov SU$(1,1)$-CS with $\varkappa=2$ and 5. These states  are {always} super-Poissonian. Left: Average value of photons $\bar n$ as a function of the parameter $u$  \eqref{avernphperel}. Right: Mandel parameter $\widetilde{Q_M}$ \eqref{persu11Mand} as a function of $\bar n$.}
	\label{fig:Perelomov_Mandel}
\end{figure}

Then, applying \eqref{QMF} with \eqref{su11bin},  i.e., $F(u) = -2\varkappa\ln(1-u)$, provides the Mandel parameter, 
\begin{equation}
\label{persu11Mand}
Q_M(u)= \frac{u}{1-u}\ \Leftrightarrow\ \widetilde{Q}_M(\bar n)=\frac{\bar n}{2\varkappa}\,.
\end{equation}
It follows that contrarily to the previous case the  states \eqref{cssu11} are super-Poissonian. This result could be expected from the above discussion about their thermal nature. The functions $\bar n(u)$ and  $\widetilde{Q_M}(\bar n)$ are plotted in Figure \ref{fig:Perelomov_Mandel} for $\varkappa=2$ and $5$. 

\begin{figure}[H]
	\centering
	\includegraphics[width=0.45\textwidth]{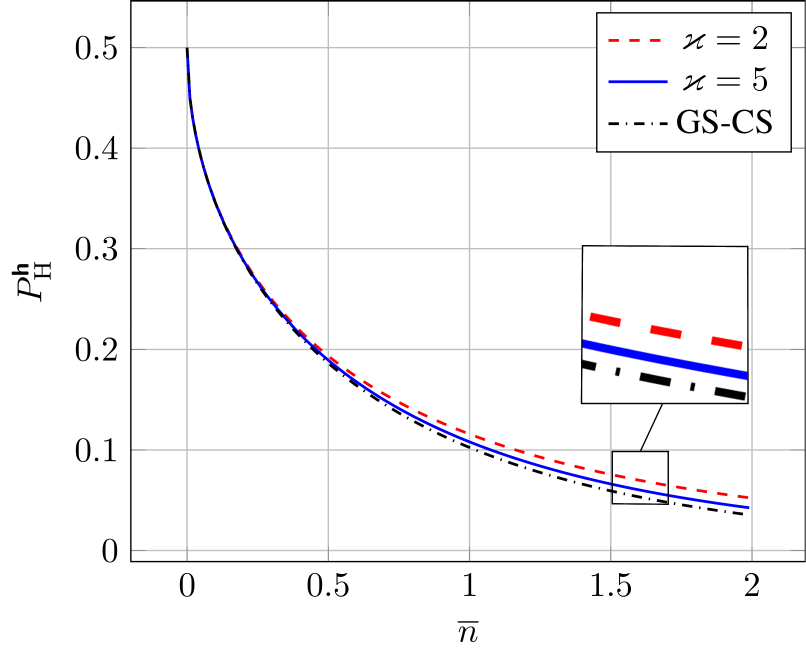}
	\includegraphics[width=0.45\textwidth]{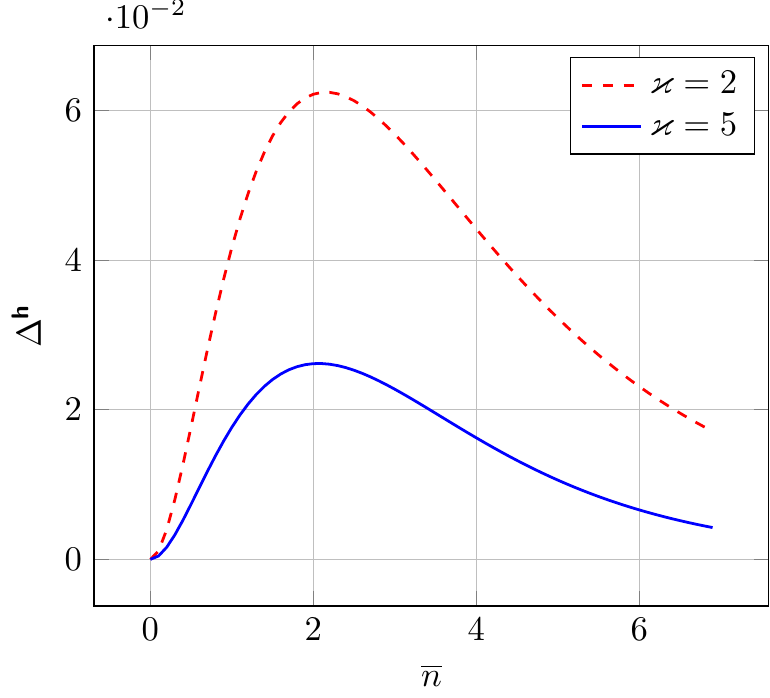}
	\caption{Helstrom bounds for GS-CS and Perelomov SU$(1,1)$-CS,  with $\varkappa=2$ and 5, are compared. This latter  states perform  { better} than the former ones. Left: The Helstrom bound $P_H^{\bsh}$ \eqref{HBAN}, for GS-CS and Perelomov SU$(1,1)$-CS. Right: The function $\Delta^{\bsh}(\bar n)$ \eqref{persu11helst}.}
	\label{fig_SHB_Perelomov}
\end{figure}

Concerning the Helstrom bound,  \eqref{Deltah} becomes here
\begin{equation}
\label{persu11helst}
\Delta(\bar n)= \left(1+\frac{\bar n}{2\varkappa}\right)^{-2\varkappa} - e^{-\bar n}\,.
\end{equation}
 Applying the  inequality \eqref{berineq} yields
$\Delta(\bar n)\geq 0$ for any $\bar n$ and $\kappa$. {Note that $\Delta(\bar n)\to 0$ as $\vk \to \infty$.} Contrarily to  the Perelomov spin $SU(2)$ CS above, the Helstrom bound is now larger than the GS-CS Helstrom bound  for any value of $u=\vert \alpha\vert^2$, this is clearly shown in Figure  \ref{fig_SHB_Perelomov}. Here again, this result could be  expected due to the ``classical'' thermal nature of the Perelomov SU$(1,1)$ coherent states.  

\subsubsection{Barut-Girardello CS}
\begin{figure}[H]
	\centering
	\includegraphics[width=0.45\textwidth]{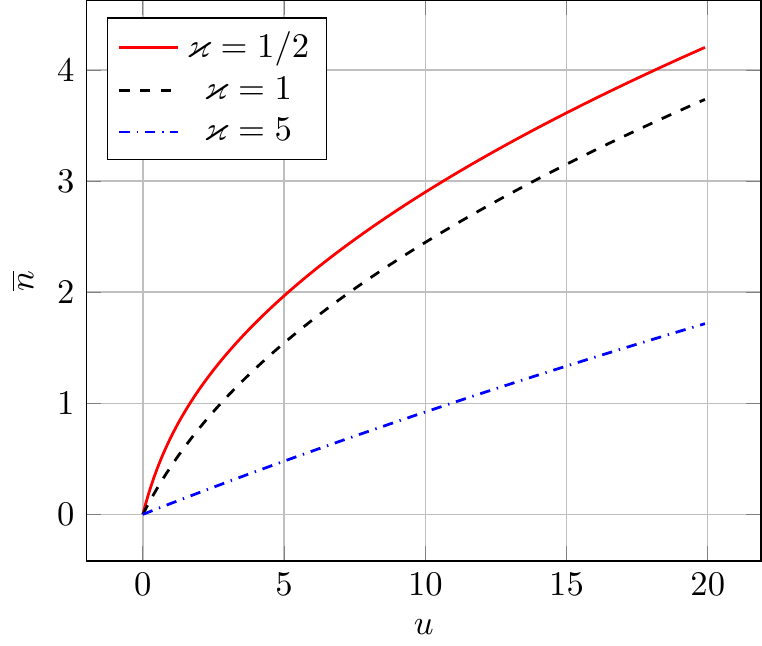}
	\includegraphics[width=0.45\textwidth]{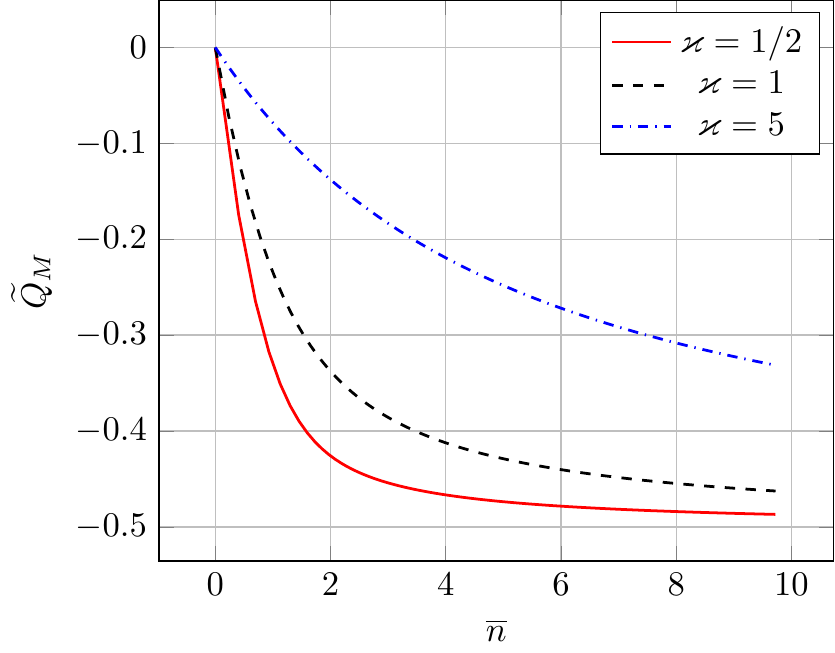}
	\caption{Photon statistics for Barut-Girardello SU$(1,1)$ CS with $\varkappa=1/2, 2$ and $5$. Left: Average value of photons $\bar n$ as a function of the parameter $u$.  Right figure: The Mandel parameter $\widetilde{Q_M}$ as a function of $\bar n$ is always negative, which makes the present states sub-Poissonian. }
	\label{fig:Barut_Mandel}
\end{figure}

{This set of nonlinear CS \cite{bargir71,angaklamopen01} belongs to the AN class. These are  eigenstates of  the lowering operator of SU$(1,1)$ in the discrete series representation $U^{\varkappa}$, with $\varkappa> 1/2$. The related Fock space $\mathcal{H}$ is also infinite whereas  $\alpha$  is allowed to assume any value  in $\C$. 
}
\begin{equation}
\label{bargirCS}
|\alpha;\varkappa\rg_{\mathrm{BG}} = \sum_{n=0}^{\infty}\alpha^n\,h_{n}\left(\vert\alpha\vert^2\right)\,|n\rg\, , \quad h_{n}(u)= \frac{ \lambda_n}{\sqrt{\mathcal{N}_{\mathrm{BG}}(u)}}\,, \quad \lambda_n= \sqrt{\frac{\Gamma(2\varkappa)}{n!\Gamma(2\varkappa + n)}}=\frac{1}{\sqrt{x_n!}}\, , 
\end{equation}
with $x_n= n(2\varkappa +n-1)$, and 
\begin{equation}
\label{ BGbessel}
\mathcal{N}_{\mathrm{BG}}(u)= \Gamma(2\varkappa)\sum_{n=0}^{\infty}\frac{u^n}{n!\Gamma(2\varkappa +n)}= \Gamma(2\varkappa)\,u^{1/2-\varkappa}\,I_{2\varkappa-1}(2\sqrt{u})\,.
\end{equation}
Here $I_{\nu}$ is a  modified Bessel function of a first kind \cite{magnus66},
\begin{equation}
\label{bess1kind}
I_{\nu}(z) = \left(\frac{z}{2}\right)^{\nu}\sum_{n=0}^{\infty}\frac{\left(\dfrac{z^2}{4}\right)^n}{n !\, \Gamma(\nu +n +1)}\,.
\end{equation}
{In this case the solution to the moment problem \eqref{mompbxn} reads as}
\begin{equation}
\label{momBG}
\int_{0}^{\infty}\ud u\,w_{\mathrm{BG}}(u)\,\frac{u^n}{\mathcal{N}_{\mathrm{BG}}(u)} = x_n!\,, \quad w_{\mathrm{BG}}(u)= \mathcal{N}_{\mathrm{BG}}(u)\,\frac{2}{\Gamma(2\varkappa)}\,u^{\varkappa-1/2}\,K_{2\varkappa-1}(2\sqrt{u})\, ,
\end{equation}
{with $K_\nu$ being the second modified Bessel function. This naturally leads to  the resolution of the identity:}
\begin{equation}
\label{BGresunit}
\int_{\C} \frac{\ud^2 \alpha}{\pi}\,w_{\mathrm{BG}}\left(|\alpha|^2\right)\, |\alpha;\varkappa\rg_{\mathrm{BG}}{}_{\mathrm{BG}}\lg\alpha;\varkappa| =\id\,. 
\end{equation}
The associated mean values \eqref{nlcsavern} and  \eqref{nclsavern2}, and the Mandel parameter have been  calculated  in \cite{angaklamopen01}, together with their asymptotic behavior at large $u$. They read respectively:
 \begin{equation}
\label{BGbarn}
\bar n (u) = \sqrt{u}\frac{I_{2\varkappa}(2\sqrt{u})}{I_{2\varkappa-1}(2\sqrt{u})} \underset{\mathrm{large} \,u}{\approx} 
\sqrt{u}-\varkappa + \frac{1}{4} + \mathrm{O}\left(\frac{1}{\sqrt{u}}\right) \quad \mbox{and} \quad \underset{\mathrm{small} \,u}{\approx}  \frac{u}{2 \varkappa}   \left(1-\frac{u}{2 \varkappa \, (1+ 2 \varkappa)}\right)  \, , 
\end{equation}
  \begin{equation}
\label{BGbarn2}
\overline{n^2} (u) = \bar n (u) + u\frac{I_{2\varkappa+1}(2\sqrt{u})}{I_{2\varkappa-1}(2\sqrt{u})} \underset{\mathrm{large} \,u}{\approx} 
u + (1-2 \varkappa) \sqrt{u} \quad \underset{\mathrm{small} \,u}{\approx} \frac{u}{2\varkappa}  \left(  1- 
\frac{(1-2 \varkappa) \, u}{2 \varkappa \, (1+2 \varkappa)}
\right)   \, , 
\end{equation}
 \begin{equation}
\label{BGsu11Mand}
Q_M(u)=  \sqrt{u}\left[\frac{I_{2\varkappa+1}(2\sqrt{u})}{I_{2\varkappa}(2\sqrt{u})} - \frac{I_{2\varkappa}(2\sqrt{u})}{I_{2\varkappa-1}(2\sqrt{u})}\right] \underset{\mathrm{large} \,u}{\approx} -\frac{1}{2} + 
\frac{1}{16} (-3 + 16 \varkappa-16 \varkappa^2) \sqrt{\frac{1}{u}}
\quad \mbox{and} \quad \underset{\mathrm{small} \,u}{\approx} -\frac{u}{2 \varkappa \, (1+2 \varkappa)} \, .
\end{equation}
Thus, the asymptotic behavior at large $u$ of the mean square error reads $(\Delta n)^2\approx \sqrt{u}/2$. Moreover, it follows from the inequality $\left(I_{\nu +1}(x)\right)^2\geq I_{\nu }(x)\, I_{\nu +2}(x)$ for all $x\geq 0$ that contrarily to the previous Perelomov SU$(1,1)$ case, the  states \eqref{bargirCS} are sub-Poissonian.
The function $\bar n(u)$ and the Mandel parameter for the Barut-Girardello SU$(1,1)$ CS, with parameters $\varkappa=1/2, 2, 5$, are shown in Figure \ref{fig:Barut_Mandel}.

\begin{figure}[H]
	\centering
	\includegraphics[width=0.45\textwidth]{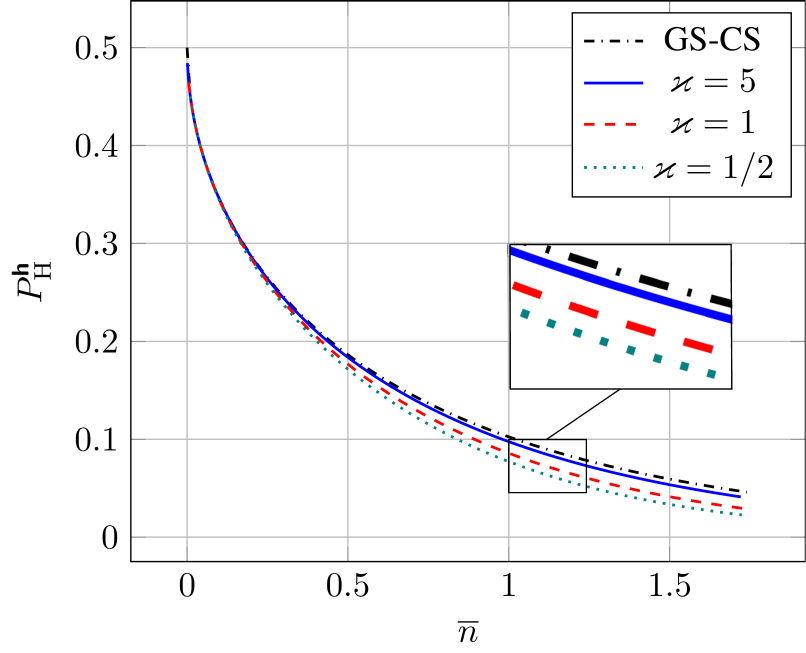}
	\includegraphics[width=0.45\textwidth]{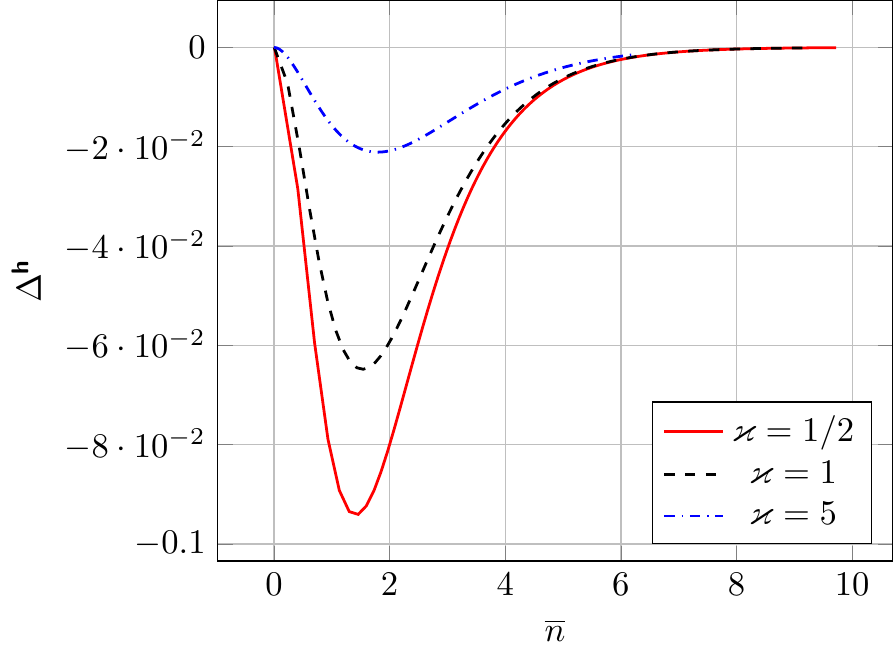}
	\caption{Left: The Helstrom bounds   
	for GS-CS and Barut-Girardello SU$(1,1)$ CS, with $\varkappa=1/2, 2$ and $5$,  are compared. 
	Right: Delta function (\ref{BGsu11helst1}). We notice that 
	the difference of the Helstrom bounds increases as $\varkappa$ decreases. 
		}
	\label{fig_SHB_BarutGirardello}
\end{figure}

As to the Helstrom bound,  \eqref{Deltah} becomes here
\begin{equation}
\label{BGsu11helst}
\Delta(\bar n)= \left[\Gamma(2\varkappa)\,(u(\bar n))^{1/2-\varkappa}\,I_{2\varkappa-1}\left(2\sqrt{u(\bar n)}\right)\right]^{-1}- e^{-\bar n}=\frac{    \widetilde{ \Delta}(\bar{n})   }{e^{\bar{n} }\,\Gamma(2\varkappa)\,(u(\bar n))^{1/2-\varkappa}\,I_{2\varkappa-1}\left(2\sqrt{u(\bar n)}\right)}\,,
\end{equation}
where
\begin{equation}
\label{BGsu11helst1}  \widetilde{ \Delta}(\bar{n})    := e^{\bar n}-\Gamma(2\varkappa)\,(u(\bar n))^{1/2-\varkappa}\,I_{2\varkappa-1}(2\sqrt{u(\bar n)})\,. \end{equation}
Helstrom bounds ($P_H^{\bsh}(\bar n)$ and $\Delta^{\bsh}(\bar n)$) for GS-CS and Barut-Girardello SU$(1,1)$ CS  are compared in Figure  \ref{fig_SHB_BarutGirardello}. Like for the spin CS, we notice that the Helstrom bound is lower than the GS-CS Helstrom bound  for any value of $u=\vert \alpha\vert^2$. 


\section{NL-CS from deformed binomial distributions}
\label{defbincs}

We now turn our attention to an important subclass of the non-linear coherent states \eqref{nlcs} associated with a sequence $\boldsymbol{\chi}$ of positive numbers $x_n$ for $n>0$ and $x_0 = 0$. Their  construction is based on the deformation of  the binomial distribution which was  used in the Bernouilli transform \eqref{probnideal} in order to take into account the imperfection of the detection encoded by the parameter $\eta$. The rationale behind such deformations was to keep the property 
\eqref{probnidealcs} enjoyed by the GS-CS's, for which imperfect detection amounts to replace $\alpha$ by  $\sqrt{\eta}\alpha$ \eqref{etaalpha}, while the deformed binomial expression is a well defined probability distribution. This means that we examine this replacement as an alternative to the change $\bar n \mapsto \eta \bar n$ in \eqref{averneeta} -- note that the latter is not equivalent to \eqref{etaalpha}.
This aim was at the origin of the series of works \cite{cugaro10},  \cite{cugaro11},  \cite{bercugaro12},  \cite{bercugaro13},  where two of us were involved. We distinguish between asymmetric and symmetric deformations of the binomial distribution.

\subsection{Asymmetric deformations}
From the sequence $\boldsymbol{\chi}=(x_0, x_1, \dotsc, x_n, \dotsc)$, $x_0 = 0$ and $x_n>0$ for $n>0$, we build the formal normalised distribution
\begin{equation}
\label{berndef}
 \mathfrak{p}_k^{(n)}(\eta) = \frac{x_n!}{x_{n-k}! \, x_k!}\, \eta^k \, p_{n-k}(\eta)\, ,
 \quad \sum_{k=0}^n \mathfrak{p}_k^{(n)}(\eta)=1\, .
\end{equation}
The polynomials $p_{s}(\eta)$  satisfy $p_s(0) = 1$, $p_s(1) = 0$, for all $s$, and the sequence $(p_{s})_{n\in \N}$ is determined by recurrence: $ p_0(\eta) = 1\,, \ p_1(\eta) = 1- \eta,\dotsc$
 If $  p_{m}(\eta) \geq 0$ for all $m$ and $\eta\in [0,1]$, then a probabilistic interpretation is valid, and $\eta= \langle x_k \rangle_n / x_n $. 
A noticeable outcome is that 
 \begin{equation}
\label{genfunction}
G_{\mathcal{N},\eta}(u):=\frac{\mathcal{N}(u)}{\mathcal{ N}(\eta u)} = \sum_{s=0}^{\infty} p_s(\eta) \frac{u^s}{x_s!}\,,  \quad \mathcal{N}(u)= \sum_n \frac{u^n}{x_n!}\,,
\end{equation}
{is the generating function of the polynomials  $p_s(\eta)$. This is important in order to  find explicitly all sequences $\boldsymbol{\chi}$ for which the Bernouilli-like distribution \eqref{berndef} obeys the positiveness condition, i.e., it is a genuine probability distribution.
As a consequence, using an alphabet based on the NL-CS \eqref{nlcs}, the probability
$\sfP_n( u; \eta)$ to detect $n$-photons with sub-unit efficiency, $\eta < 1$, is given in terms of the perfect, $\eta = 1$, detector probability
}
 \eqref{poissondef}, 
$\sfP_m( u; \eta =1) =  u^m/(\mathcal{N}( u) x_m!) $, $ u = \vert \alpha \vert^2$, by the deformed Bernoulli transform:
 \begin{equation}
\label{probnidealnla}
\sfP_n( u; \eta) = \sum_{m=n}^{\infty} \frac{x_m!}{x_{m-n}! \, x_n!}\, \eta^n \, p_{m-n}(\eta)\,\sfP_m( u;\eta = 1)= \frac{(\eta\, u )^n}{\mathcal{N}(\eta  u)\, x_n!} \,.
\end{equation}
This discrete probability distribution corresponds to the normalized states $ |\sqrt \eta\alpha;\boldsymbol{\chi}\rg$, being equivalent to the family of non-linear coherent states defined by \eqref{nlcs} through the rescaling \eqref{etaalpha}.

{One can interpret such deformation as a modification of the probability of getting $k$ wins (photon clicks) in a total of $n$ events. With $y_0= 0$, $y_n = x_n/n$ for $n>0$ $
\mathfrak{p}_k^{(n)}(\eta)$
 gives the probability  to get $k$ wins and $n-k$ losses, while 
\begin{equation}
\label{probpi}
  \pi_{k}^{(n)}(\eta) \deq \frac{\mathfrak{p}_k^{(n)}(\eta)}{\binom{n}{k}}= \dfrac{y_n!}{y_{n-k}! \, y_k!}\,\eta^k \, p_{n-k}(\eta) 
\end{equation}
 stands for the probability of one given string having $k$ wins and $n-k$ losses. 
Such a deformation means there are correlations between different events.
 We call this distribution asymmetric because $\mathfrak{p}_k^{(n)}(\eta)$ is not invariant under  $k \to n-k$ and $\eta \to 1-\eta$, as it would hold  for the binomial case. Due to this asymmetry, there might be bias in favor of either win or loss, including when $\eta = 1/2$. 
 }

{ A comprehensive analysis exploring generating functions to examine the positiveness condition $p_{m}(\eta) \geq 0$ for all $m$ and $\eta\in [0,1]$ and finding the solutions was exposed in \cite{bercugaro12}.
It starts with $\mathcal{N}(u)$ which gives the generating function of  $p_k(\eta)$  through the relation (\ref{genfunction}). Let  $\Sigma$ be the set of entire series $\mathcal{N}(u)=\sum_{n=0}^\infty a_n u^n$ having a strictly positive radius of convergence and verifying the conditions $a_0=1$ and $\forall n \ge1, \, a_n>0$. Let us  associate with  $\mathcal{N}$ the  sequence  $x_n=a_{n-1}/a_n=n \, \mathcal{N}^{(n-1)}(0)/\mathcal{N}^{(n)}(0)$ so that  $a_n=1/(x_n!)$.
  Next  let $\Sigma_0$ be the set of entire series $\sum_{n=0}^\infty a_n z^n$ having a strictly positive radius and verifying $a_0=0$, $a_1 >0$ and $\forall n \ge 2, \, a_n \ge 0$. 
It was proved in \cite{bercugaro12} that 
$
\Sigma_+:=\left\{ \mathcal{N} \in \Sigma \, | \,  \forall \eta \in [0,1), \, p_n(\eta) > 0 \, \right\}=\{ e^F \, | \, F \in \Sigma_0\}
$
 is the set of deformed exponentials with generating functions $G_{\mathcal{N},\eta}(u)$ that do solve the positiveness problem. All sequences $\boldsymbol{\chi}$ with this probabilistic content  are called \textit{sequences of complete statistical type}, or CST.
 }
 
{ 
One can add to this probabilistic interpretation of the above defined asymmetric deformed binomial distribution the Poisson-like limit behaviour. 
Indeed, if the sequence $ \{ x_n\}_{n\in \mathbb{N}}$ is such that, , at fixed $m$,
$\underset{n \to \infty}{\lim}{\frac{x_{n-m}}{x_n}} = 1$, then
the limit when  $n \to \infty$ of  $ \mathfrak{p}_k^{(n)}(\eta)$, given by \eqref{berndef},  with  $\eta =  u/x_n$ is equal to $ \frac{1}{\mathcal{N}(u)}\, \frac{ u^k}{x_k!}$. 
 }
 
{ This allows to better comprehend the behaviour of CST sequences in comparison with natural integers.
In fact, when picking a  $\mathcal{N} \in \Sigma_+$; its related CST sequence} $x_n = n \, \mathcal{N}^{(n-1)}(0)/\mathcal{N}^{(n)}(0)$ verifies  $ 0 \le x_n \le n x_1.$
 We give now two examples (see \cite{bercugaro12} for more).

 \subsubsection{Example 1} 
{
 Let  $\{a_n \}_{n \in \mathbb{N}}$ be a sequence of positive real numbers with $\sum_{n=0}^\infty a_n < \infty $. Given $\zeta \in [0,1]$ the function $
\mathcal{N}(u)=\prod_{k=0}^\infty (1+ \zeta a_k u)/(1-a_k u)
$
belongs to $\Sigma_+$, see \cite{bercugaro12}. One gets  a simple illustration with 
$\mathcal{N}(u) = (1-a u)^{-m} = \sum_{k=0}^\infty (u^k)/(x_k^{(m)}!) \,,
$
where $a>0$. The convergence radius is $1/a$, and the the resulting sequence is  equivalent, up to a variable rescaling, to the Perelemov SU$(1,1)$ case \eqref{su11bin}, 
$x_k^{(m)}! = k! \,(m-1)!/a^k(m-1+k)!,$ and  $x_k^{(m)} = k/(a (m-1+k).$ 
It is worth noting that at the limit $k \rightarrow \infty$ all coefficients $x_k^{(m)}$ go to the limit $x_\infty^{(m)} = 1/a$. 
Moreover, the resulting polynomial probabilities $p_k(\eta)$, $0 \leq \eta \leq 1$,  are hypergeometrical
$
p_k(m;\eta) = {}_2F_1(-m,-k;1-k-m;\eta)  .
$
For example, with $n=2$,  the probability of having two wins in a set of two trials is similar to 
the binomial case, yet the probability to get 
one win and one loss is smaller and the probability to get two losses is higher in comparison to the binomial one. Nonetheless the sum of all three possibilities is equal to one, as required. 
}

\subsubsection{Example 2} 
\begin{figure}[htb]
	\centering
	\includegraphics[width=0.45\textwidth]{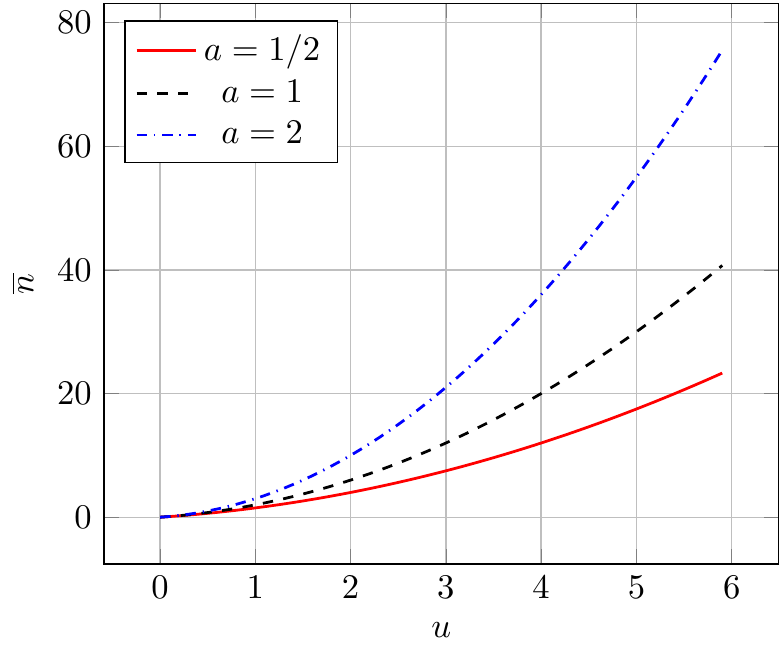}
	\includegraphics[width=0.45\textwidth]{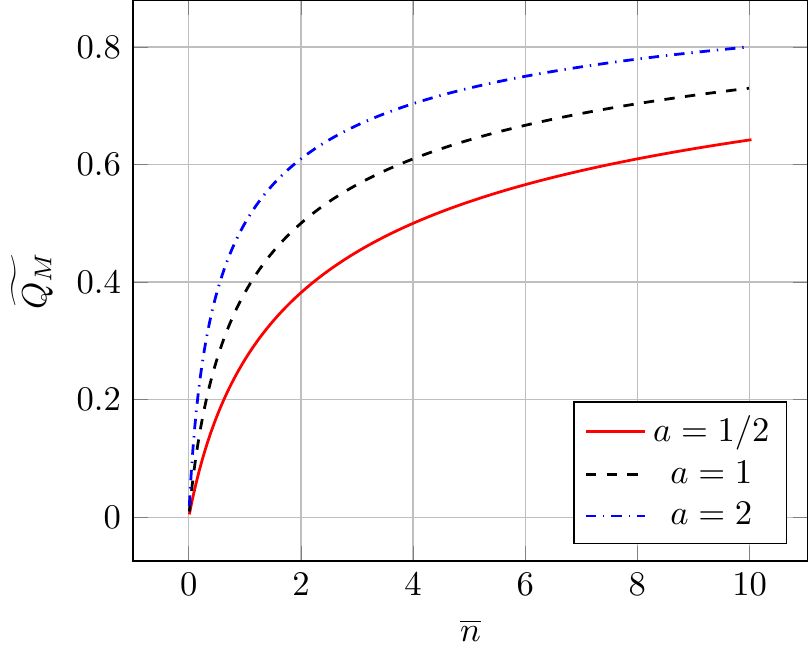}
	\caption{Photon statistics for Hermite polynomials CS with $a=1/2, 1$ and $2$. 
	Left: $\bar{n}$ as function of $u$. 
	Right: 
Mandel parameter versus $\bar{n}$. We notice that it is positive, which makes these states super-Poissonian, as is the case for all NL-CS based on deformed binomial distributions, see Sec. \ref{demo}.	}
	\label{fig:Hermite_Mandel}
\end{figure}

{Another example is yielded by the  function 
}
\begin{equation}
\label{NuHermite}
\mathcal{N}(u) = \exp \left(u + \frac{a}{2} u^2\right)\, \quad a>0\,.
\end{equation}                  
{The convergence radius is infinite, as for the exponential function.
The associated generating function reads}
$G_{\mathcal{N},\eta}(u)= 
\exp \left(( 1-\eta) t + \frac{a}{2} (1-\eta^2) u^2      \right) \,.
$
The corresponding $x_n!$ are expressed in terms of Hermite polynomials:
\begin{equation}
\label{xnfat2} 
x_n! = \left[  \frac{\ii^n \left(\frac{a}{2}\right)^{n/2}}{n!} H_n \left( 
\frac{-\ii}{\sqrt{2 a}}
\right) \right]^{-1} = \left\lbrack\sum_{m=0}^{\left\lfloor\frac{n}{2}\right\rfloor} \frac{(a/2)^m}{m!(n-2m)!} \right\rbrack^{-1}\,.
\end{equation}
and we have the asymptotic behavior 
$x_n \sim \sqrt{n/a} + \cdots \,.
$
\\
The average average value of number of photons is given by	
\begin{equation}
\bar n(u)=u(1+au)
\quad\Leftrightarrow \quad
u(\bar n)=\frac{1}{2a}\left( \sqrt{1+4a\bar n}-1\right) \, .
\end{equation}
The Mandel parameter is
\begin{equation}
Q_M(u)=\frac{au}{1+au}
\quad\Leftrightarrow \quad
\widetilde{Q_M}(\bar n)=\frac{\sqrt{1+4a\bar n}-1}{\sqrt{1+4a\bar n}+1}\,,
\end{equation}
and is positive for all possible values of $\bar n$, as shown in the Fig. \ref{fig:Hermite_Mandel}. The related states are thus all super-Poissonian, as is the case for all NL-CS based on deformed binomial distributions, see Sec. \ref{demo}.

\begin{figure}[htb]
	\centering
	\includegraphics[width=0.45\textwidth]{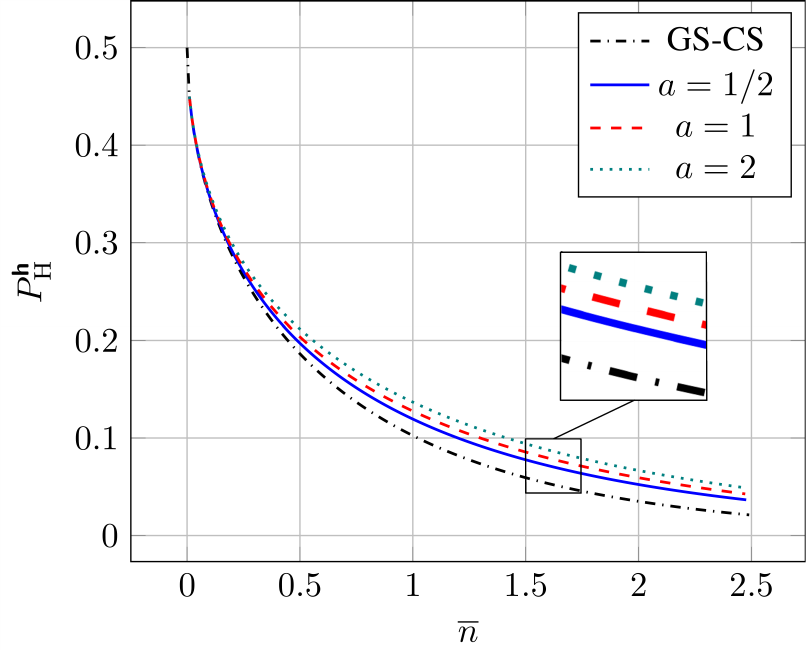}
	\includegraphics[width=0.45\textwidth]{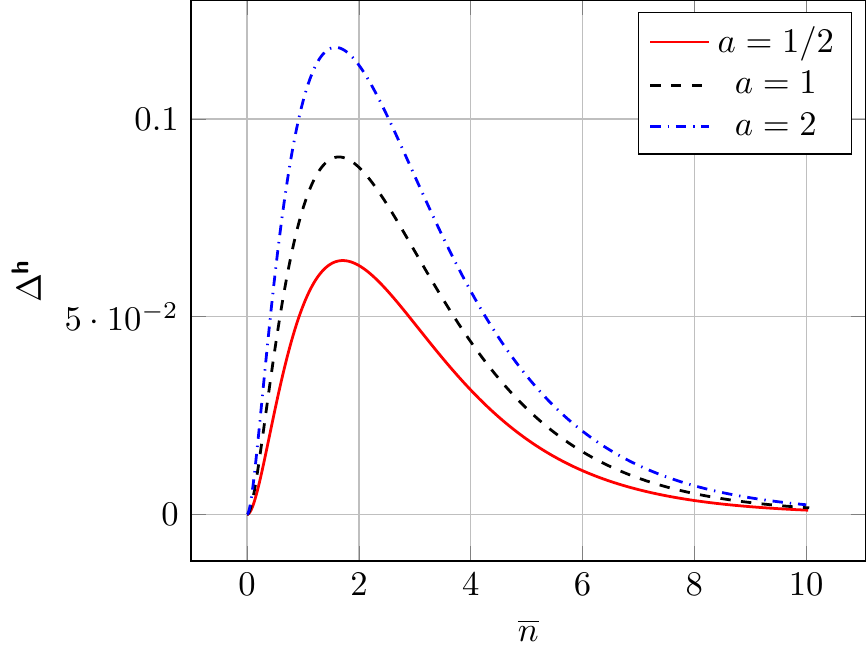}
	\caption{Helstrom bounds for GS-CS and Hermite polynomials CS, with $a=1/2, 1$ and $2$, are compared. 
	Left: Helstrom bounds as function of $\bar{n}$ for several values of $a$.  
	The Helstrom bound is higher than GS-CS,  
	which means that the quantum error limit is larger than for GS-CS states, as is the case for all NL-CS based on deformed binomial distributions, see Sec. \ref{demo}. Right: Delta function (\ref{delta_herm}) versus $\bar{n}$. }
	\label{fig_SHB_Hermite}
\end{figure}

The function $\Delta^{\bsh}(\bar n)$ given by \eqref{PhHPh} reads
\begin{equation}
\label{delta_herm}
\Delta(\bar n)=e^{\frac{1}{4a}-\frac{\bar n}{2}-\frac{1}{4a}\sqrt{1+4a\bar n}}-e^{-\bar n}\,,
\end{equation}
and is always positive, that is the Helstrom bound is higher than for GS-CS, see Fig. \ref{fig_SHB_Hermite}.

\subsection{Symmetric deformations}
\label{sec_symmetric}

The following symmetric option was explored in \cite{bercugaro13}:
\begin{equation}
\label{dist_symmetric}
\mathfrak{p}_k^{(n)}(\eta)=\dfrac{x_n!}{x_{n-k}! x_k!} q_k(\eta) q_{n-k}(1-\eta)\\,
\end{equation}
{ with $q_k(\eta)$ being polynomials of  degree $k$. 
Normalization and non-negativeness conditions constrain the  $\mathfrak{p}_k^{(n)}(\eta)$ through }
\begin{equation}
\label{ }
\begin{split}
&\sum_{k=0}^n \mathfrak{p}_k^{(n)}(\eta)=1\quad \forall n \in \mathbb{N}, \quad \forall \eta \in [0,1]\,,  \\
 &\mathfrak{p}_k^{(n)}(\eta) \ge 0\quad \forall n,k \in \mathbb{N}, \quad \forall \eta \in [0,1]\,.
\end{split}
\end{equation}
 The normalization  implies
$ \mathfrak{p}_0^{(0)}(\eta)=q_0(\eta) q_0(1-\eta)=1 \Rightarrow  q_0(\eta)=\pm 1$, $ \forall \eta \in [0,1]$.
Assuming $q_0(\eta)=1$ we have 
$
 \mathfrak{p}_0^{(n)}(\eta)=q_n(1-\eta).
$
%
%
%
{Also the positiveness condition corresponds to the positiveness of the polynomials $q_n$, for $\eta\in[0,1]$.
Like for the asymmetric case,  the polynomials $\mathfrak{p}_k^{(n)}(\eta)$  might be interpreted as the probability to get $k$ wins and $n-k$ failures in a string of $n$ trials.  In both cases there is correlation between different trials. However, since the  invariance under  $k \to n-k$ and $\eta \to 1-\eta$ holds as for  the binomial distribution, no bias can exist favoring either win or failure for $\eta = 1/2$. 
}

{An appealing example of  generating functions gives rise to binomial-type polynomials.
Let us pick  a $x_n$-generating function $\mathcal{N}(u) \in \Sigma$. Then we have
\begin{equation}
\label{gensym}
\forall \eta \in [0,1], \quad \tilde G_{\mathcal{N},\eta}(u):= \mathcal{N}(u)^\eta=\sum_{n=0}^\infty \dfrac{q_n(\eta)}{x_n!} u^n.
\end{equation}
 }
 
 {
Because $q_0(\eta)=1$ and $\forall n\ne 0, q_n(0)=0$, the positiveness condition implies that the function $\mathcal{N}(u)^\eta$ belongs to $\Sigma$. Furthermore,  It was proven in \cite{bercugaro13} that if $\mathcal{N}(u)\in \Sigma_+$, then  the polynomials $\mathfrak{p}_k^{(n)}(\eta)$ allow a  probabilitic interpretation.
Moreover one finds that the expectation value of the $k$ variable takes the simple form $\langle k \rangle_n = \eta n\, ,$ as in the usual binomial case. 
}

{A first illustration of symmetric deformations} is found from the Perelomov SU$(1,1)$ type \eqref{su11bin} (up to a rescaling of the variable) with
$
\mathcal{N} (u) = \left( 1- u/s \right)^{-s}\, , \quad s > 0\, . 
$
 This gives the sequence  $
x_n = n s /(n + s -1)$ with $ \lim_{n\to \infty}x_n = s.$ 
The corresponding polynomials are  
$
q_n(\eta) =   (s \eta)_n/ (s)_n, 
$ where $(x)_n = \Gamma(x+n)/\Gamma(x)$ { (Pochhammer symbol).  
 In addition to the mean value expression $\langle k \rangle_n = \eta n$, the square root of the variance
$
(\sigma_k)_n = n \sqrt{\eta(1-\eta)}\sqrt{(1 + s/n)/(1 + s)} 
$
 also becomes  proportional to  the mean value at large $n$. 
}

\subsubsection{Example: {The modified Abel polynomials}}
\label{sec:example2}
\begin{equation}
\label{genex2}
\NN (t) =e^{-\beta W(-u/\beta)}\, , \quad \beta > 0\, , 
\end{equation}
{with $W$ being the Lambert function. The latter is defined implicitly through $W(u)e^{W(u)} = u$. 
It is worth noting that if $\beta \to \infty$ thus $\NN(u) \to e^u$. This leads to a sequence which is bounded by $\beta/e$:
}
\begin{equation}
\label{xnex2}
x_n = \frac{n\beta}{n + \beta }\left(1- \frac{1}{n + \beta}\right)^{n-2}\, , \quad \lim_{n\to \infty}x_n = \beta/e \, . 
\end{equation}
{One can also notice that $x_n \to n$ as $\beta \to \infty$.  This results into the following factorials:}
\begin{equation}
\label{factabel}
x_n!= n!\, \frac{\beta^{n-1}}{(n+\beta)^{n-1}}\, .
\end{equation}
The  polynomials $q_n$'s {read}
\begin{equation}
\label{xnex22}
q_n(\eta) =  \eta \frac{\left(\eta + \frac{n}{\beta}\right)^{n-1}}{\left(1 + \frac{n}{\beta}\right)^{n-1}}\, . 
\end{equation}
{Note that  $q_0(\eta) = 1$ and $q_1(\eta) = \eta$. 
The probability distribution }
\eqref{dist_symmetric} reads in this case
\begin{equation}
\label{abelprobdist}
\mathfrak{p}_k^{(n)}(\eta) = \binom{n}{k}\eta(1-\eta) \frac{(\eta + k/\beta)^{k-1}(1-\eta + (n-k)/\beta)^{n-k-1}}{(1+n/\beta)^{n-1}}\,. 
\end{equation}

\begin{figure}[H]
	\centering
	\includegraphics[width=0.45\textwidth]{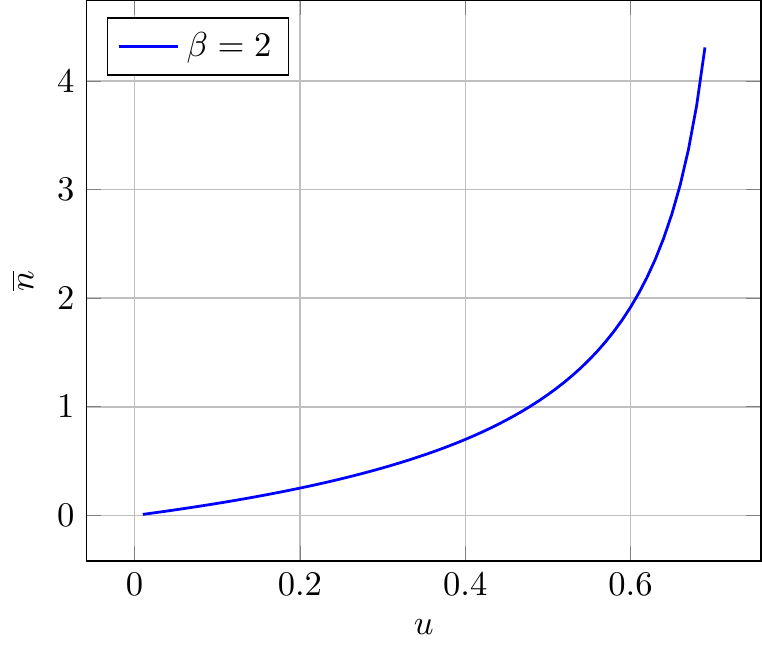}
	\includegraphics[width=0.45\textwidth]{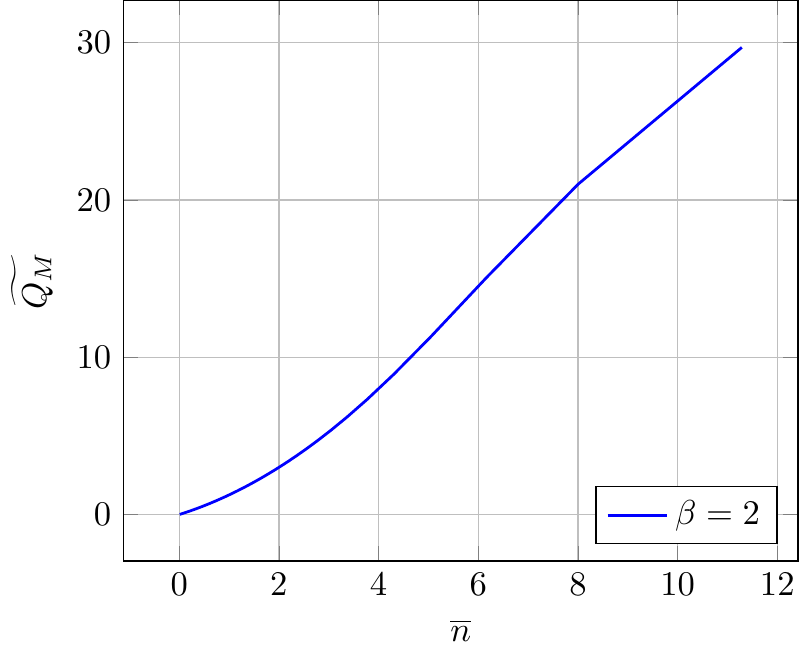}
	\caption{Photon statistics for modified Abel polynomials CS with $\beta=2$. Left: $\bar{n}$ as function of $u$. 
	Right: Mandel parameter versus $\bar{n}$. 
	We notice that it is positive, which makes these states super-Poissonian, as is the case for all NL-CS based on deformed binomial distributions, see Sec. \ref{demo}.	}
	\label{fig:Abel_Mandel}
\end{figure}

\begin{figure}[H]
	\centering
	\includegraphics[width=0.45\textwidth]{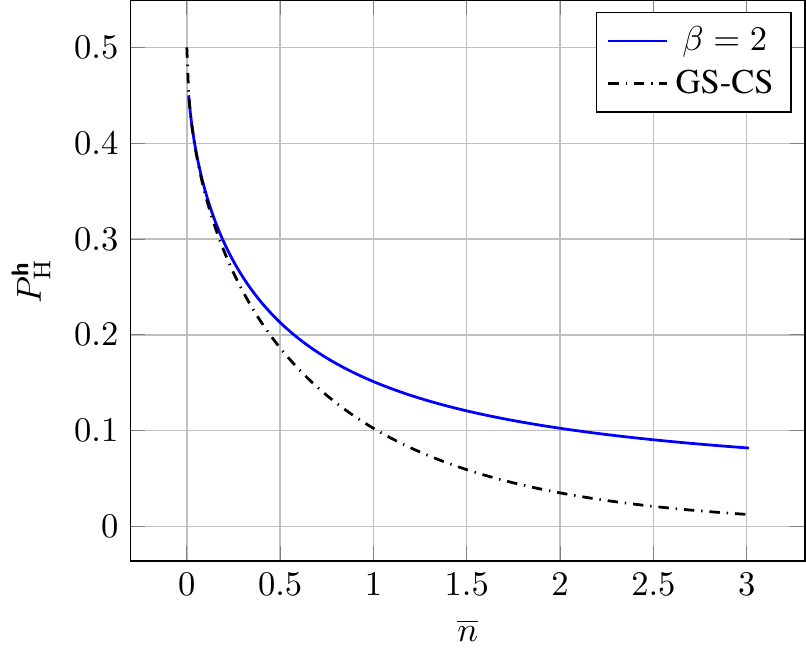}
	\includegraphics[width=0.45\textwidth]{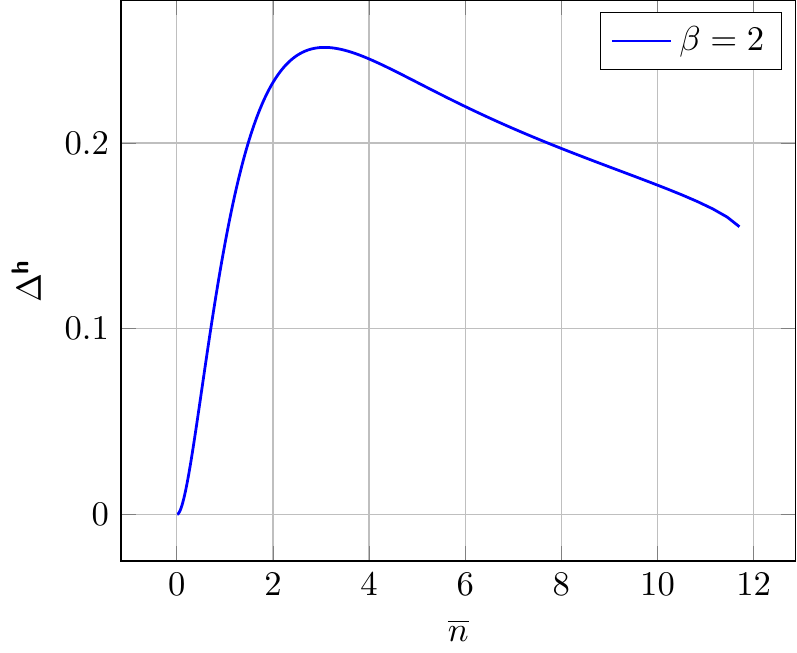}
	\caption{Helstrom bounds for GS-CS and modified Abel polynomials CS with $\beta=2$ are compared. Left: Helstrom bounds as function of $\bar{n}$. Right: Delta function (\ref{delta_herm}) versus $\bar{n}$.
	The Delta function is always positive as is the case for all NL-CS based on deformed binomial distributions, see Sec. \ref{demo}.	}
	\label{fig_SHB_Abel}
\end{figure}

The average value of photons and Mandel parameter are then given by
\begin{equation}
\bar n(u)= -\frac{\beta W \left(-\frac{u}{\beta} \right)}{1+ W\left(-\frac{u}{\beta} \right)}\,, 
\qquad 
Q_M(u)=\frac{1}{\left( 1+ W\left(-\frac{u}{\beta} \right)\right) ^{2}}-1\, .
\end{equation}
The inversion $u(\bar n)$ is done here numerically in order to get the Mandel parameter and Helstrom bound as a function of $\bar n$. These states are super-Poissonian (see Fig. \ref{fig:Abel_Mandel}), and lead to a Helstrom bound higher than the GS-CS case (see Fig. \ref{fig_SHB_Abel}).

\subsection{Quasi-classical nature of CS associated with asymmetric and symmetric binomial deformations} 
\label{demo}
We now show that the above results, namely states are super-Poissonian and HB is higher compared to GS-CS, hold for any symmetric or asymmetric deformation of the binomial law. Let us consider a generic deformed exponential defined through
\begin{equation}
\label{genexp2}
\mathcal{N}(u) = \sum_{n=0}^{\infty}\frac{u^n}{x_n!},
\end{equation}
and demonstrate that this always yields positive Mandel parameter and $\Delta(\bar n)$.
\beprop
Symmetric and asymmetric binomial deformations defined through \eqref{berndef} and \eqref{dist_symmetric}, respectively, lead to super-Poissonian NL-CS. 
\enprop
\bprf
This can easily be shown by considering the sequence $\lbrace x_n\rbrace$ generated by $\mathcal{N}(u)\in\Sigma_+$.
Expressing $\mathcal{N}$ as $\mathcal{N}(u)=e^{F(u)}\in\Sigma_+$, with $F(u) = \sum_{k=1}^\infty a_k u^k$, 
we have 
\begin{equation}
\label{formand1}
\overline{n^2}-\overline{n}^2-\overline{n}=\sum_{k=1}^{\infty}(k^2-k) a_k u^k\, .
\end{equation}
Since $\forall n\geq 2$ $a_n\geq0$, the right-hand side of \eqref{formand1} is positive. Consequently the Mandel parameter $Q_M$, defined through \eqref{mandelQu}, is positive. 
\eprf

\beprop 
If $\mathcal{N}(u)\in\Sigma_+$, the NL-CS cannot optimize the Helstrom bound
\enprop
\bprf
In \cite{bercugaro12} one can establish the proof from the following statements:
\begin{itemize}
	\item For all $\mathcal{N}\in\Sigma_+$, $\ln\mathcal{N}\in\Sigma_0$ and
	\begin{equation}
	\label{eqcond2}
	\ln\mathcal{N}(u)=\sum_{k=1}^{\infty}\frac{(-p'_k(1))}{x_k!}\frac{u^k}{k}\, ,
	\end{equation}
	
	\item $p'_1(1)=-1$ and $p'_n(1)\leq 0$\, .
\end{itemize}
Using \eqref{eqcond2} we get
\begin{equation}
\label{eqcond31}
u\frac{d}{du}\ln\mathcal{N}(u)= \sum_{k=1}^{\infty}\frac{(-p'_k(1))}{x_k!}u^k\, .
\end{equation}
Since $u>0$,  we conclude that 
\begin{equation}
\ln\mathcal{N}( u)<u\frac{\ud}{\ud u}\ln\mathcal{N}(u)\, ,
\end{equation}
which can be written as
\begin{equation}
\label{eqfr33}
\frac{1}{\mathcal{N}( u)}>e^{-u\frac{\ud}{\ud u}\ln\mathcal{N}(u)}\, . 
\end{equation}
The expression above, once expressed in terms of $\bar n$, leads to
\begin{equation}
\label{eqfr22}
\left|h_0(u(\overline{n})) \right|^2>e^{-\overline{n}}\, . 
\end{equation}
Therefore, the $\Delta_h$ function is always positive and therefore the Helstrom bound is larger than in the GS-CS case.
\eprf

\section{Mandel parameter and Helstrom bound for both Susskind-Glogower  and  modified Susskind-Glogower CS}
\label{sussglo}
Here we  examine the Susskind-Glogower CS's \cite{sussglog64} which exhibit attractive properties on different levels, as shown in  \cite{montelmoya11,montelmoyasoto11,moyasoto11}. On the mathematical side, they are built through the action of the unitary displacement operator on the Fock vacuum state, see for instance  \cite{regomopa08}. 
\begin{equation}
\label{VVD}
\R \ni x \mapsto |x\rg_{\mathrm{SG}} = D_{\mathrm{SG}}(x)|0\rg\equiv  e^{x (V^{\dag}- V)}|0\rg\,.
\end{equation}
The operator $V$ and its adjoint $V^{\dag}$ are the shift operators in the Fock space:
\begin{equation}
\label{defCCR}
V = \sum_{n=0}^{\infty}|n\rg\lg n+1|\, , \quad V^{\dag} = \sum_{n=0}^{\infty}|n+1\rg\lg  n|\, ,\quad [V,V^{\dag}]=\id -|0\rg\lg0|\,. 
\end{equation}
The normalised states \eqref{VVD} expand in terms of number states as:  
\begin{equation}
\label{SGmoya}
|x\rg_{\mathrm{SG}}= \sum_{n=0}^{\infty}  (n+1)\,\frac{J_{n+1}(2x)}{x}\,|n\rg\,, 
\end{equation}
{where  $J_\nu$ is the Bessel function,}
\begin{equation}
\label{bessel}
J_\nu(z)= \left(\frac{z}{2}\right)^{\nu}\,\sum_{m=0}^{\infty}\frac{(-1)^m \left(\frac{z}{2}\right)^{2m}}{m!\,\Gamma(\nu+m+1)}\, . 
\end{equation} 
Normalisation implies the interesting identity whose proof is given  in the appendix of \cite{montelmoyasoto11}: 
\begin{equation}
\label{normbessel}
\sum_{n=1}^{\infty}  n^2\,\left(J_{n}(2x)\right)^2 = x^2\, .
\end{equation}
On the physical side, the authors of \cite{montelmoyasoto11} showed that the coherent states \eqref{SGmoya} may be modeled by propagating light in semi-infinite arrays of optical fibers.

\begin{figure}[H]
	\centering
	\includegraphics[width=0.45\textwidth]{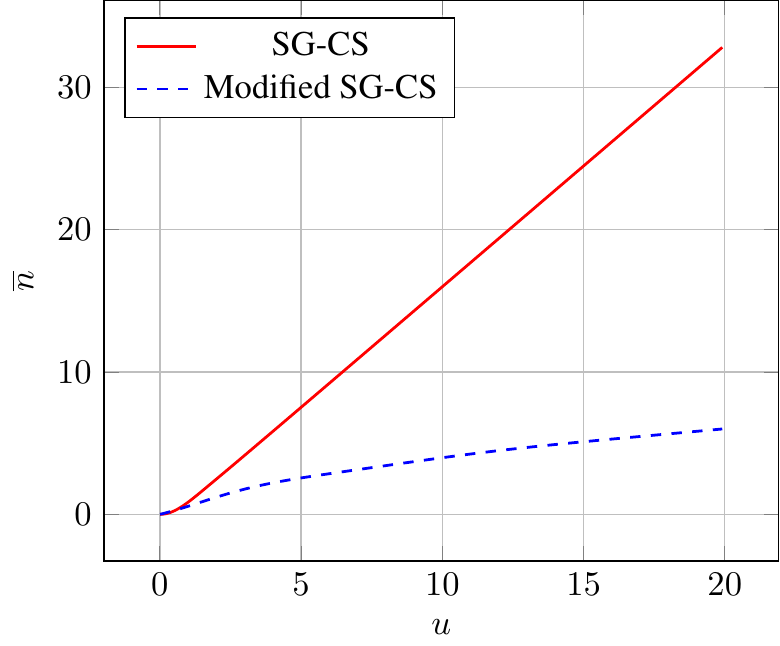}
	\includegraphics[width=0.45\textwidth]{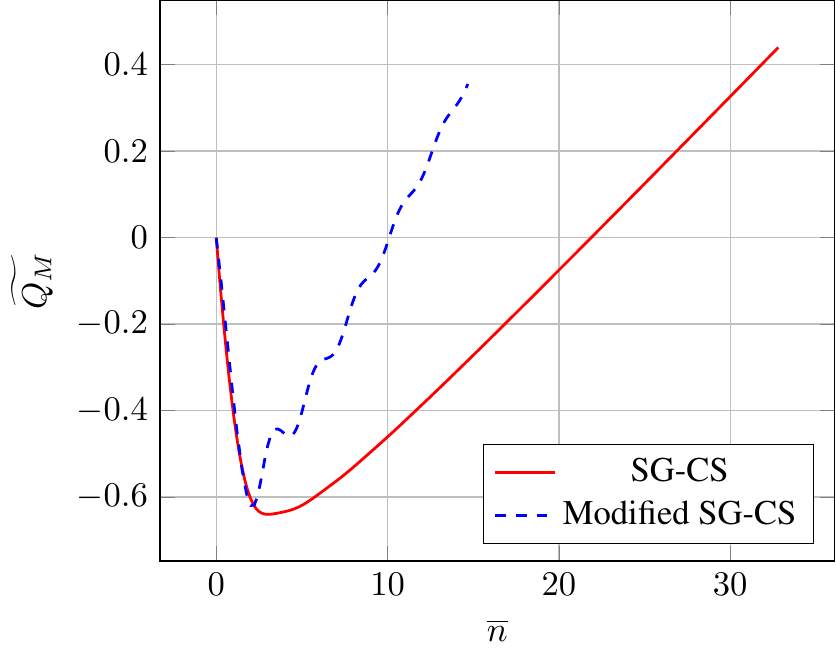}
	\caption{Photon statistics for SG-CS and modified SGm-CS. Left: $\bar{n}$ as function of $u$. 
	Right: Mandel parameter versus $\bar{n}$. 
	We notice that the SG-CS and SGm-CS are sub-Poissonian for $\bar n \lesssim 22$ and $\bar n \lesssim 10$ respectively.
	}
	\label{fig:SG_mandel}
\end{figure}

{Now, the  expression \eqref{bessel} can be used to extend \eqref{SGmoya} in a non analytic way  to a complex $\alpha$ as}
\begin{equation}
\label{extendRSG}
(n+1)\frac{J_{n+1}(2x)}{x} \mapsto \alpha^n \,(n+1)\sum_{m=0}^{\infty}\frac{(-1)^m \vert \alpha\vert^{2m}}{m!\,\Gamma(n+m+2)}\equiv \alpha^n\,h^{\mathrm{SG}}_n(\vert\alpha\vert^2)\,,  
\end{equation}
i.e., \begin{equation}
\label{vpnbessel}
h^{\mathrm{SG}}_n(u) = (n+1)\,\frac{1}{u^{\frac{n+1}{2}}}\,J_{n+1}(2\sqrt{u})\, , 
\end{equation}
and thus
\begin{equation}
\label{CSSG}
|\alpha\rg_{\mathrm{SG}}= \sum_{n=0}^{\infty} \alpha^n\, h^{\mathrm{SG}}_n(\vert\alpha\vert^2)\,|n\rg\,. 
\end{equation}
The moment equation \eqref{condortho} reads here 
\begin{equation}
\label{SGmom}
\int_{0}^{\infty}\ud u\,\frac{w(u)}{u}\,\left(J_{n}(2\sqrt{u})\right)^2 = 2\int_{0}^{\infty}\ud t\,\frac{w(t^2)}{t}\,\left(J_{n}(2t)\right)^2 = \frac{1}{n^2}\,, 
\end{equation}
and appear not to have an immediate solution. On the other hand, 
by examining the following integral formula for Bessel functions 
\cite{magnus66}, 
\begin{equation}
\label{intbessel2}
\int_0^{\infty}\frac{\ud t}{t} \, \left(J_{n}(2 t)\right)^2= \frac{1}{2n}\, .  
\end{equation}
we are  led to replace the SG-CS of \eqref{SGmoya} by the modified \textit{Susskind-Glogower coherent states} (SGm-CS):
\begin{equation}
\label{SGm}
|\alpha\rg_{\mathrm{SGm}}= \sum_{n=0}^{\infty}\alpha^n\,h^{\mathrm{SGm}}_n(\vert\alpha\vert^2) \,|n\rg\,, \quad h^{\mathrm{SGm}}_n(u) = \sqrt{\frac{n+1}{\mathcal{N}(u)}}\,\frac{1}{u^{\frac{n+1}{2}}}\,J_{n+1}(2\sqrt{u})\, ,
\end{equation}
with 
\begin{equation}
\label{normSGm} 
\mathcal{N}(u) = \frac{1}{u}\sum_{n=1}^{\infty}  n\,\left(J_{n}(2\sqrt{u})\right)^2=\frac{1}{u}\left[ 2u\left(J_0(2\sqrt{u})\right)^2- \sqrt{u} J_0(2\sqrt{u}) J_1(2\sqrt{u})+ 2u\left(J_1(2\sqrt{u})\right)^2\right] \,.
\end{equation}
{The formula \eqref{intbessel2} then can be used to demonstrate that the resolution of the identity \eqref{resid} is satisfied for SGm-CS $|\alpha\rg_{\mathrm{SGm}}$  with $w(u)= \mathcal{N}(u)$.}
 \begin{equation}
\label{residSGm}
 \int_{\C}\frac{\ud^2\alpha}{\pi}\,\mathcal{N}\left(\vert\alpha\vert^2\right)\, |\alpha\rg_{\mathrm{SGm}}{}_{\mathrm{SGm}}\lg\alpha| =\id\, .
\end{equation}
Mean values \eqref{nlcsavern} and  \eqref{nclsavern2} read respectively:
 \begin{equation}
\label{SGmbarn}
\begin{aligned}
\bar n(u)=\frac{1}{\mathcal{N}(u)}-1\,,
\end{aligned}
\end{equation}
\begin{equation}
\label{SGmbarn2}
\begin{aligned}
\overline{n^2}(u)= -\frac{2}{\mathcal{N}(u)} + \frac{4}{3}(2u+1)-\frac{u\left(J_0(2\sqrt{u})\right)^2-u\left(J_1(2\sqrt{u})\right)^2+\sqrt{u}J_0(2\sqrt{u})J_1(2\sqrt{u})}{3u\mathcal{N}(u)}\,.
\end{aligned}
\end{equation}
For comparison, Mandel parameters for SG-CS and SGm-CS are plotted in  Figure \ref{fig:SG_mandel}.
It results that the SG-CS and SGm-CS are sub-Poissonian for $\bar n \lesssim 22$ and $\bar n \lesssim 10$) respectively.

\begin{figure}[H]
	\centering
	\includegraphics[width=0.45\textwidth]{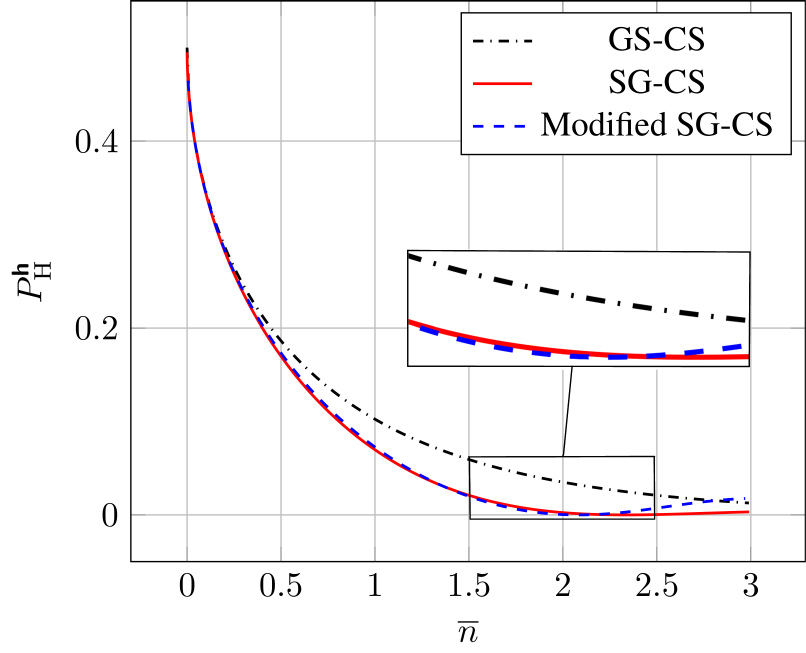}
	\includegraphics[width=0.45\textwidth]{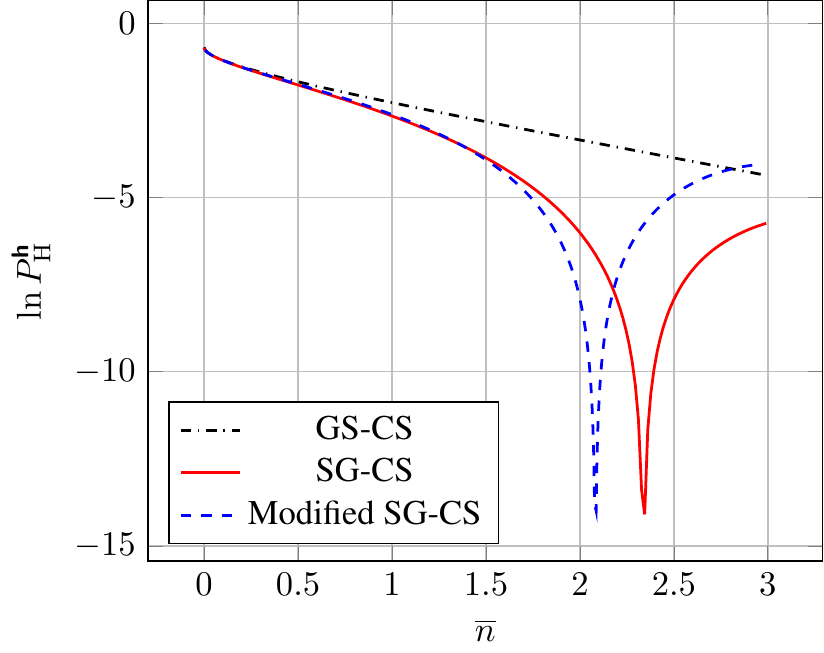}
	\caption{Helstrom bounds for GS-CS, SG-CS and modified SG-CS are compared. Left figure: Helstrom bound $P_H^{\bsh}$ \eqref{HBAN}, for GS-CS, SG-CS and modified SGm-CS. Right figure: Natural logarithm of the Helstrom bound $\ln P_H^{\bsh}$, for GS-CS, SG-CS and  SGm-CS, which makes it easier to see for which values of $\bar n$ the Helstrom bounds reaches zero.}
	\label{fig_SHB_SG}
\end{figure}

As to the Helstrom bound, \eqref{Deltah} becomes
\begin{equation}
\label{SGmhelst}
\begin{aligned}
\Delta(\bar n(u))= \frac{\left(J_{1}(2\sqrt{u})\right)^2}{u\mathcal{N}(u)}
-e^{1-1/\mathcal{N}(u)}\,.
\end{aligned}
\end{equation}
This is compared to both SG and GS coherent states in Figure \ref{fig_SHB_SG}. 
Astonishingly the Helstrom bounds vanishes for some specific values of $\bar n$ in the two cases of GS-CS and GSm-CS. Taking the logarithm makes it easier to see for which values of $\bar n$ the Helstrom bounds reaches zero, this happens for $\bar{n}$ around two in both cases.

\begin{figure}[H]
	\centering
	\includegraphics[width=0.45\textwidth]{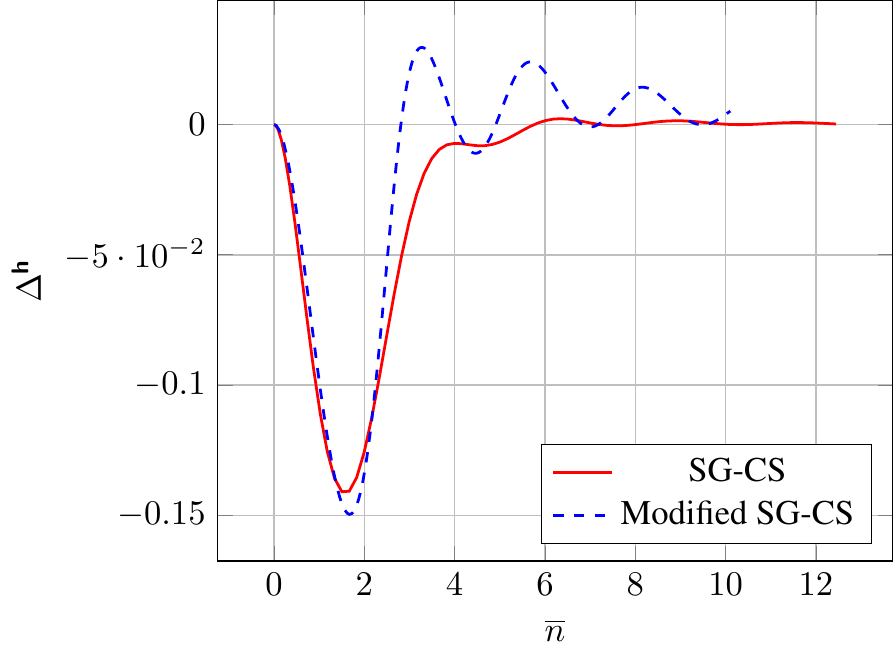}
	\caption{Function $\Delta^{\bsh}(\bar n)$ for SG-CS and modified SG-CS. The Helstrom bounds for GS-CS and GSm-CS oscillate as $\bar n$ is varying and, actually, do also vanish for higher values of $\bar n$. Yet the difference with GS-CS gets lower and so does $\Delta^{\bsh}(\bar n)$. }
	\label{}
\end{figure}

\section{Conclusion}
\label{conclu}

In this work we have examined statistical properties of various generalizations of the Glauber-Sudarshan coherent states within the framework of Quantum Optics, namely Fock spaces of one-mode photons. In particular we analysed the Mandel parameter and more importantly the quantum error limit, or Helstrom bound.
We have proved the classical or quasi-classical  nature of a whole family of non-linear coherent states, those based on symmetric and asymmetric deformations of the binomial distribution, in the sense that they are super-Poissonian and do not improve the GS-CS Helstrom bound. 

We have also revealed the deep quantum nature of four other families of NL-CS, namely the Perelomov Spin CS's, the SU$(1,1)$ Barut-Girardello CS's, the Susskind-Glogower CS's, and  their modified version: they are sub-Poissonian  and improve more or less significantly the GS-CS Helstrom bound. In particular, the SG-CS's and SGm-CS's turn out to exhibit an astonishing result: the Helstrom bound vanishes for some values of the average photons number $\bar n$, the smallest being around $\bar n=2$. This means that these states are then orthogonal to the vacuum state. Consequently it appears possible to use coherent states for binary communication with no quantum error, nevertheless, limited to specific values of $\bar n$.

In a parent work we will focus our attention on the modified  Susskind-Glogower CS. These, unlike the SG-CS, obeys the normalisation \eqref{condortho}, which in turn implies the resolution of the identity in the Fock space \eqref{resid}, or \eqref{residSGm} in the present case. Finally, we will  explore their physical feasibility and relevance in the spirit of \cite{montelmoyasoto11}.



\section*{Acknowledgements}
The authors acknowledge CNPq, FAPERJ and CAPES for financial support. JPG thanks CBPF for 
hospitality and PCI support.


\begin{thebibliography}{99}

\bibitem[Holevo 1973] {holevo73} A.~S. Holevo, Statistical decision theory for quantum systems, \textit{J. Multivariate Anal.} \textbf{3} (1973) 337394.


\bibitem[Holevo 2011] {holevo11} A.~S. Holevo, 
\newblock \textit{Probabilistic and statistical aspects of quantum theory},
\newblock Quaderni/Monographs, 1. Edizioni della Normale, Pisa, 2011. 

\bibitem[Fuchs 1996]{main:ch4:fuchsphd} C.~A. Fuchs,
\newblock \textit{ Distinguishability and Accessible Information in
Quantum Theory},
\newblock Ph.D. thesis, University of New Mexico, 1996.

\bibitem[Peres 1995]{main:peres} A. Peres,
\newblock \textit{ Quantum Theory: Concepts and Methods}.
\newblock Kluwer Academic Publishers, Dordrecht, 1995.

 \bibitem[Bennet\&Brassard 1984]{main:ch4:benbra} C.~H. Bennett and G. Brassard, 
\newblock in \textit{ Proceedings of the
IEEE International Conference on Computers, Systems
abd Signal Processing}
\newblock IEEE, New York, 1984, 175.

\bibitem[Helstrom 1976]{main:helstrom} C.~W. Helstrom.
\newblock \textit{ Quantum Detection and Estimation Theory}.
\newblock Academic Press, New York, 1976.

\bibitem[Cook et al 2007]{main:ch4:comage} R.~L. Cook, P.~J. Martin, and J.~M. Geremia, Optical coherent state discriminationusing a closed-loop quantum measurement,  \textit{Nature},  \textbf{446} (2007) 774.

\bibitem[Tsujino et al 2011]{tsujino11} K. Tsujino, D. Fukuda, G. Fujii, S. Inoue, M. Fujiwara, M. Takeoka, and M. Sasaki, Quantum receiver beyond the standard quantum limit of coherent optical communication, \textit{Phys. Rev. Lett.}, \textbf{106} (2011) 250503.

\bibitem[Becerra et al 2013]{becerra13} F.~ E. Becerra, J. Fan, G. Baumgartner, J. Goldhar, J.~ T. Kosloski and A. Migdall, Experimental demonstration of a receiver beating the standard quantum limit for multiple nonorthogonal state discrimination, \textit{Nat. Photonics}, \textbf{7(2)} (2013) 147.

\bibitem[Kunz et al  2019]{Kunz:2019}  L.~Kunz  \textit{et al.},  Beating the Standard Quantum Limit for Binary Constellations in the Presence of Phase Noise, 21st International Conference on Transparent Optical Networks (ICTON) (2019).
 
\bibitem[Sych \& Leuchs 2016]{Sych:2016} D.~Sych and G.~Leuchs,  Practical Receiver for Optimal Discrimination of Binary Coherent Signals, \textit{Phys. Rev. Lett.} \textbf{117} (2016) 200501.

\bibitem[DiMario et al 2019]{DiMario:2019} M.~T.  DiMario,  \textit{et al.}, Optimized Communication Strategies with Binary Coherent States over Phase Noise Channels,  \textit{npj Quantum Information} \textbf{5} (2019) 65. 

\bibitem[Gazeau 2009]{gazeaubook09} J.-P. Gazeau, 
\newblock \textit{Coherent States in Quantum Physics}. 
 \newblock Wiley-VCH, Berlin, 2009.

\bibitem[Perina 1984]{perina} J. Peri\~{n}a
\newblock \textit{Quantum Statistics of Linear and Nonlinear Optical Phenomena}.
\newblock Reidel, Dordrecht, 1984.

\bibitem[Perina 1994]{perina01} J. Peri\~{n}a (Editor).
\newblock \textit{ Coherence and Statistics of Photons and Atoms}.
\newblock Wiley-Interscience, 2001.

\bibitem[Tse et al 2019]{Tse:2019wcy} M.~Tse \textit{et al.}  Quantum-Enhanced Advanced LIGO Detectors in the Era of Gravitational-Wave Astronomy, \textit{Phys. Rev. Lett.} \textbf{123} (2019)  231107.

\bibitem[Acernese et al 2019]{Acernese:2019sbr} F.~Acernese \textit{et al.}, Increasing the Astrophysical Reach of the Advanced Virgo Detector via the Application of Squeezed Vacuum States of Light, \textit{Phys. Rev. Lett.}  \textbf{123} (2019)  231108.

\bibitem[Gazeau 2019]{gazeau19} J.-P. Gazeau, Coherent states in Quantum Optics:  An oriented overview, \textit{Integrability, Supersymmetry and Coherent States, A volume in honour of  V. Hussin, Eds. Kuru, Sengul, Negro, Javier, Nieto, Luis M., 
CRM series in Mathematical Physics}. Springer, (2019) 69-101;


\bibitem[Curado et al 2010]{cugaro10} E.~M.~F. Curado, J.-P. Gazeau, and L.~M~.C.~S. Rodrigues, Non-linear coherent states for optimizing Quantum Information, \textit{Proceedings of the Workshop on Quantum Nonstationary Systems,  October 2009,  Brasilia. Comment section (CAMOP)} \textit{Phys. Scr.}, \textbf{82} (2010) 038108.

 \bibitem[Curado et al 2012]{cugaro11} E.~M.~F. Curado, J.-P. Gazeau, and Ligia M.~ C.~ S. Rodrigues, On a Generalization of the Binomial Distribution and Its Poisson-like Limit, \textit{J. Stat. Phys.}, \textbf{146} (2012) 264-280.

\bibitem[Bergeron et al 2012]{bercugaro12} H. Bergeron, E.~M.~F. Curado, J.-P. Gazeau, and Ligia M.~ C.~ S. Rodrigues, Generating functions for generalized binomial distributions, \textit{J. Math. Phys.},  \textbf{53} (2012) 103304-1-22.

\bibitem[Bergeron et al 2013]{bercugaro13} H. Bergeron, E.~M.~F. Curado, J.-P. Gazeau, and Ligia M.~ C.~ S. Rodrigues, Symmetric generalized binomial distributions, \textit{J. Math. Phys.},  \textbf{54} (2013) 123301-1-22.

\bibitem[Moya-Cessa and Soto-Eguibar 2011]{moyasoto11} H.~M. Moya-Cessa and F. Soto-Eguibar, \textit{Introduction to Quantum Optics}. Rinton Press, Paramus, 2011. 

\bibitem[Loudon 1973]{main:loudon} R. Loudon. 
\newblock \textit{ The Quantum Theory of Light}.
\newblock Oxford Univ. Press,
Oxford, 1973.

\bibitem[Geremia 2004]{geremia04} J.~M. Geremia,
Distinguishing between optical coherent states with imperfect detection, \textit{Phys. Rev. A} \textbf{70} (2004) 062303-1-9.

\bibitem[Akhiezer 1965]{akhiezer65} N.~I. Akhiezer,  \textit{The classical moment problem and some related questions in analysis}, translated from the Russian by N. Kemmer, Hafner Publishing Co., New York, 1965.

\bibitem[Perelomov 1972]{perel72} A.~M. Perelomov,  Coherent States for Arbitrary Lie Group, \textit{Commun. math. Phys.}, \textbf{26} (1972) 222-236.  

\bibitem[Perelomov 1986]{perel86} A.~M. Perelomov, \textit{Generalized Coherent States and Their Applications}. Springer, Berlin (1986).

\bibitem[Fox 2006]{fox06} M. Fox, \textit{Quantum Optics: An Introduction}. Oxford University Press, New York (2006).

\bibitem[Ali et al 2008]{algahel08} S.~T. Ali, J.-P. Gazeau, and B. Heller,  Coherent states and Bayesian duality, \textit{ J. Phys. A: Math. Theor.}, \textbf{41} (2008) 365302-1-22. 

%
%

\bibitem[Wodkiewicz-Eberly 1985]{wodkiewicz85} {K. Wodkiewicz and J. Eberly,  Coherent states, squeezed fluctuations, and the SU$(2)$ and SU$(1, 1)$ groups in quantum-optics applications. \textit{J. Opt. Soc.  B}, \textbf{2} (1985) 458.}

\bibitem[Gerry 1987]{gerry87} {C. C. Gerry,  Application of SU$(1, 1)$ coherent states to the interaction of squeezed light
in an anharmonic oscillator, \textit{Phys. Rev. A}, \textbf{35} (1987) 2146.}

\bibitem[Brif 1995]{brif95} {C. Brif, Photon states associated with the Holstein-Primakoff realization of the SU$(1,1)$ Lie algebra
\textit{Quantum Semiclass. Opt.}, \textbf{7}  (1995)  803.}

\bibitem[Hach et al 2016]{hachetal16} E. E. Hach III, P. M. Alsing, and C. C. Gerry, Violations of a Bell in- equality for entangled SU$(1,1)$ coherent states based on dichotomic observables, \textit{Phys. Rev. A}, \textbf{93} (2016) 042104.

\bibitem[Hach et al 2018]{hachetal18} E. E. Hach III, R. BirritellaI, P. M. Alsing, and C. C. Gerry,  SU$(1,1)$ parity and strong violations of a Bell inequality by entangled Barut-Girardello coherent states, \textit{J. Opt. Soc.  B}, \textbf{35} (2018) 2433.

\bibitem[Aharonov et al 2011]{aharonov73} Y. Aharonov, E.~C. Lerner, H.~W. Huang, and J.~M. Knight, Oscillator phase states, thermal equilibrium and group representations, \textit{J. Math. Phys.}, \textbf{14} (2011) 746-755.

\bibitem[Barut-Girardello 1971]{bargir71} A.~O. Barut and L. Girardello, New ``Coherent'' States Associated with Non-Compact Groups, \textit{Commun. Math. Phys.}, \textbf{21} (1971) 41-55.

\bibitem[Antoine et al 2001]{angaklamopen01} J.-P. Antoine, J.-P. Gazeau, J.~R. Klauder,  P. Monceau, and K.~A. Penson, \textit{J. Math. Phys.}, \textbf{42} (2001) 2349-2387.

\bibitem[Magnus et al 1966]{magnus66}    W. Magnus, F. Oberhettinger, and R.~P.  Soni.
\newblock \textit{ Formulas and Theorems for
the Special Functions of Mathematical Physics}.
 \newblock Springer-Verlag,  Berlin, Heidelberg and New York (1966).




 \bibitem[Susskind-Glogower 1964]{sussglog64} L. Susskind and J. Glogower, Quantum mechanical phase and time operator,
 \textit{Phys. Phys. Fiz. 1}, \textbf{1} (1964) 49-61.
 
\bibitem[Montel-Moya-Cessa 2011]{montelmoya11} R. De J. Le\'on Montel and H.~M. Moya-Cessa, Modeling non-linear coherent states in fiber arrays,  \textit{Int. J.  Quant. Information},  \textbf{9} (2011) Suppl. 349-355.

\bibitem[Montel et al 2011]{montelmoyasoto11} R. De J. Le\'on Montel,  H.~M. Moya-Cessa and F. Soto-Eguibar, Nonlinear coherent states for the Susskind-Glogower operators, \textit{Rev. Mex. F\'{\i}s.}, \textbf{57} (2011) 133-147.


\bibitem[R\'ecamier et al 2008]{regomopa08} J. R\'ecamier, M. Gorayeb, W.~L. Moch\'an, and J.~L. Paz, Nonlinear Coherent States and Some of Their Properties, \textit{Int. J.
Theor. Phys.}, \textbf{47} (2008) 673-683.

\end{thebibliography}
\end{document}